\documentclass[12pt,preprint]{aastex}

\usepackage{color}
\usepackage{epsfig}
\usepackage[fleqn]{amsmath}
\usepackage{multirow}
\usepackage{multicol}
\usepackage{mathptmx}
\usepackage{graphicx}
\usepackage{bm}
\usepackage{mathrsfs}

\shorttitle{Dynamics of charged particles} \shortauthors{Sun et
al.}

\begin{document}


\title{\large Dynamics of charged particles moving around Kerr black
hole with inductive charge and external magnetic field }

\author{Xin Sun$^{1}$, Xin Wu$^{1,2,3, \dag}$, Yu Wang$^{1}$, Chen Deng$^{1}$,  Baorong Liu$^{1,3}$, Enwei Liang$^{1,3}$}
\affil{1. School of Physical Science and Technology, Guangxi
University, Nanning 530004, China \\
2. School of Mathematics, Physics and Statistics $\&$
Center of Application and Research of Computational Physics, Shanghai
University of Engineering Science, Shanghai 201620, China \\
3. Guangxi Key Laboratory for Relativistic Astrophysics, Guangxi
University, Nanning 530004, China} \email{$\dag$ Corresponding
Author's Email: wuxin$\_$1134@sina.com, xinwu@gxu.edu.cn 
}

\begin{abstract}

We mainly focus on the effects of small changes of  parameters on
the dynamics of charged particles around the Kerr black hole
surrounded by an external magnetic field, which can be considered
as a tidal environment. The radial motions of charged particles on
the equatorial plane are studied via an effective potential. It is
found that the particle energies at the local maxima values of the
effective potentials increase with an increase of the black hole
spin and the particle angular momenta, but decrease with an
increase of one of the inductive charge parameter and magnetic
field parameter. The radii of stable circular orbits on the
equatorial plane also increase, whereas those of the innermost
stable circular orbits decrease. On the other hand, the effects of
small variations of the parameters on the orbital regular and
chaotic dynamics of charged particles on the non-equatorial plane
are traced by means of a time-transformed explicit symplectic
integrator, Poincar\'{e} sections and fast Lyapunov indicators. It
is shown that the dynamics sensitively depends on small variations
of the inductive charge parameter, magnetic field parameter,
energy and angular momentum. Chaos occurs easily as each of the
dynamical parameters increases. When the dragging effects of the
spacetime increase, the chaotic properties are not always weakened
under some circumstances.

\end{abstract}



\section{Introduction}

The Kerr metric that describes  a rotating black hole is a
solution of the Einstein's field equations of general relativity.
The observed event-horizon-scale images of the supermassive black
hole candidate in the center of the giant elliptical galaxy M87
are consistent with  the dark shadow of a Kerr black hole
predicted by general relativity [1]. The motion of a particle in
the vicinity of the Kerr black hole is integrable because of the
existence of four conserved quantities including the energy,
angular momentum, rest mass and azimuthal motion of the particle.
The azimuthal motion corresponds to the Carter constant [2], which
is obtained from the separation of variables in the
Hamilton-Jacobi equation.

Observational evidences demonstrate the existence of strong
magnetic fields in the vicinity of the supermassive black hole at
the centre of the Galaxy [3]. The external magnetic fields which
can be considered as a tidal environment are generally believed to
play a crucial role in the transfer of the energy from the
accretion disk to jets. Radiation reaction depending on the
external magnetic field strength causes the accretion of charged
particles from the accretion disk to shift towards the black hole.
An inductive charge introduced by Wald [4] generates an induced
electric field due to a contribution to the Faraday induction from
the parallel orientation of the spin of a black hole and the
magnetic field. When the inductive charge takes the Wald charge,
the potential difference between the horizon of a black hole and
infinity vanishes, and  the process of selective accretion is
completed [5, 6].  The effects of the magnetic fields involving
the induced electric field are so weak in comparison to the
gravitational mass effects that they do not change the spacetime
metrics. However, they can essentially affect the motion of
charged test particles in accreting matter if the ratio of the
electric charge and mass of the particle is large. In most cases,
the fourth invariable quantity related to the azimuthal motion of
the particles is absent when the external electromagnetic fields
are considered near the black hole. Thus, the dynamics of charged
test particles in the black holes with external electromagnetic
fields is nonintegrable.

Although the magnetic fields in the vicinity of the black holes
destroy the integrability of these spacetimes in many problems,
the radial motions of the charged particles on the equatorial
plane are still integrable and solvable. It is mainly studied by
means of an effective potential. The effective potential seems
simple, but it describes many important properties of the
spacetimes. In particular, unstable circular orbits, stable
circular orbits, and innermost stable circular orbits (ISCOs) on
the equatorial plane are clearly shown through the effective
potential. It is interesting to study these equatorial orbits in
the theory of accretion disks. An accreted material with
sufficient angular momentum relative to an axisymmetric massive
central body will be still attracted by the
central body, but such force will be compensated due to the large
angular momentum. This easily forms an accretion disk. However,
the accreted material without sufficient angular momentum will
fall into the central body [7-9]. Electromagnetic fields could
influence dynamics of charged particles in accreting matter,
therefore, the ISCOs in the field of a magnetized black hole are
shifted towards the horizon for a suitable spin direction. In
other words, the inner boundary of the accretion disk goes towards
the central body. In view of the importance of the topic on the
effective potential and stable circular orbits on the equatorial
plane, the topic has been taken into account in a large number of
literatures [7-28]. These problems discussed in the existing works
are based on the equatorial plane. In some extended theories of
gravity, such as Brans-Dicke gravity, scale-dependent gravity and
asymptotically safe gravity in the context of black hole physics
[29-35], the effective potentials, unstable circular orbits,
stable circular orbits and ISCOs on a plane slightly different
from the equatorial can be discussed similarly.

When the external magnetic fields destroy the spacetime's symmetry
(precisely speaking, the external magnetic fields lead to the
absence of the fourth constant related to the particles' azimuthal
motion), the generic motion of charged particles on the
non-equatorial plane can be chaotic in some circumstances. If the
external magnetic fields do not destroy the symmetry, no chaotic
dynamics is possible. For example, charged particle motions in the
Kerr-Newman black hole spacetime are regular and nonchaotic
because of the existence of four integrals leading to the
integrability of the system [36]. Chaos describes a dynamical
system sensitive dependence on initial conditions. The theory of
chaotic scattering in the combined effective potential of the
black hole and the asymptotically uniform magnetic field is useful
to explore  the mechanism hidden behind the charged particle
ejection [5].  The energy of the charged particle in such combined
fields is split into one energy mode along the magnetic field line
direction and another energy mode at the perpendicular direction.
The chaotic charged particle dynamics in the combined
gravitational and magnetic fields leads to an energy interchange
between the two energy modes of the charged particle dynamics. As
a result, it can provide sufficient energy to ultra-relativistic
motion of the charged particle along the magnetic field lines.
Based on the importance of studies of the chaotic motion in the
gravitational field of a black hole combined with an external
electromagnetic field, many authors [5, 6, 12, 20, 23, 37-46] are
interested in this field.

The detection of the chaotical behavior requires  the adopted
computational scheme with  reliable results. Without doubt,
higher-order  numerical integrators such as an eighth- and
ninth-order Runge-Kutta-Fehlberg integrator with adaptive step
sizes can yield high-precision numerical solutions. However, they
are more computationally demanding than lower-order solvers. For
Hamiltonian systems, the most appropriate solvers are symplectic
integrators which respect the symplectic nature of Hamiltonian
dynamics and show no secular drift in energy errors [47-53].
A symplectic integrator method for numerical
calculation of charged particle trajectory is well known due to
its small error in energy even for long integration times, which
make it perfectly suited for the description of regular and
chaotic dynamics through Poincar\'{e} sections calculations [54].
Because the variables are inseparable in Hamiltonian systems
associated to curved spacetimes, the standard explicit symplectic
integrators do not work when these Hamiltonian systems are
separated into two parts. In this case, completely implicit
symplectic methods including the implicit midpoint method [55, 56]
and Gauss-Runge-Kutta methods [41, 54, 57, 58] are often
considered. Explicit and implicit combined symplectic methods
[59-63] take less cost than these completely implicit methods, and
then are also used. Recently, explicit symplectic integrators were
proposed for nonrotating black holes when the Hamiltonians of
these black holes have several splitting parts with analytical
solutions as explicit functions of proper time [64-66]. With the
aid of time transformations, explicit symplectic integrators are
easily available for the Kerr type spacetimes [67].

The authors of [54] employed the Gauss-Legendre
symplectic solver (i.e., s-stage implicit symplectic Runge-Kutta
method) to study the regular and chaotic dynamics of charged
particles around the Kerr background endowed with an axisymmetric
electromagnetic test field with the aid of Poincar\'{e} sections.
The authors of [68] applied the time-transformed explicit
symplectic integrators introduced in [67] to mainly explore the
effect of the black hole spin on the chaotic motion of a charged
particle around the Kerr black hole immersed in an external
electromagnetic field. Unlike Ref. [68], the present work
particularly focuses on how a small change of  the black hole
inductive charge [6] exerts influences on the effective potential,
stable circular orbits and ISCOs on the equatorial plane, and a
transition from order to chaos of orbits on the non-equatorial
plane. The effects of other dynamical parameters such as the
magnetic field parameter are also considered. For this purpose, we
introduce a dynamical model for the description of charged
particles moving around the Kerr black hole immersed in an
external magnetic field in Sect. 2. The effective potential,
stable circular orbits and ISCOs on the equatorial plane are
discussed in Sect. 3.  The explicit symplectic integrators are
designed for this problem, and the dependence of the orbital
dynamical behavior on the parameters is shown in Sect. 4. Finally,
the main results are concluded in Sect. 5.

\section{Kerr black hole immersed in external magnetic field}

The Kerr black hole is the description of a rotating  black hole
with mass $M$ and angular momentum $a$. In the standard
Boyer-Lindquist coordinates $(t, r, \theta, \phi)$, its time-like
metric is written as $ds^{2}=-c^2d\tau^2$, that is,
\begin{eqnarray}
ds^{2} &=& g_{\alpha\beta}dx^{\alpha}dx^{\beta}=g_{tt}c^2dt^2+2g_{t\phi}cdtd\phi \nonumber \\
&& +g_{rr}dr^2 +g_{\theta\theta}d\theta^2+g_{\phi\phi} d\phi^2.
\end{eqnarray}
These nonzero components in this metric are found in the paper of
[69] as follows:
\begin{eqnarray}
g_{tt} &=& -(1-\frac{2GMr/c^2}{\Sigma}), \nonumber \\
g_{t\phi} &=& -\frac{(2GMr/c^2)a\sin^{2}\theta}{\Sigma}, \nonumber \\
g_{rr} &=& \frac{\Sigma}{\Delta}, ~~~~~~~~~~
g_{\theta\theta}=\Sigma,  \nonumber \\
g_{\phi\phi} &=& (\rho^2+\frac{2GMr}{\Sigma}
a^{2}\sin^{2}\theta)\sin^{2}\theta, \nonumber
\end{eqnarray}
where $\Sigma=r^2+a^2\cos^{2}\theta$, $\Delta=\rho^2-2GMr/c^2$ and
$\rho^2=r^2+a^2$. $\tau$ and $t$ are proper and coordinate times,
respectively. $c$ is the speed of light, and $G$ denotes the
gravitational constant.

Suppose  the Kerr black hole is immersed in an external
asymptotically uniform magnetic field, which has strength $B$ and
yields an induced charge $Q$. Set $\xi^{\alpha}_{(t)}$ and
$\xi^{\alpha}_{(\phi)}$ as time-like and space-like axial Killing
vectors.  An electromagnetic four-vector potential can be found in
Refs. [6] and [70] and is written as
\begin{equation}
A^{\alpha}=aB\xi^{\alpha}_{(t)}+\frac{B}{2}\xi^{\alpha}_{(\phi)}-\frac{Q}{2}\xi^{\alpha}_{(t)}.
\end{equation}
This potential has two nonzero covariant components
\begin{eqnarray}
A_{t} &=& g_{t\alpha}A^{\alpha}= (aB-\frac{Q}{2})g_{tt}+\frac{B}{2}g_{t\phi}, \\
A_{\phi} &=& g_{\phi\alpha}A^{\alpha}=
(aB-\frac{Q}{2})g_{t\phi}+\frac{B}{2} g_{\phi\phi}.
\end{eqnarray}
When $Q=2aB_w$, the inductive charge is the Wald charge $Q_W$, and
$B_w$ is a magnetic field corresponding to the Wald charge [4].
The induced charge like the Wald charge $Q_W$ is so small that it
has no contribution to the background geometry of the black hole
[71]. However, the induced charge can exert an important influence
on the motion of a charged particle under some circumstances, as
will be shown in later discussions.

The motion of the particle around the rotating black hole embedded
in the external magnetic field is described by the Hamiltonian
\begin{eqnarray}
H &=& \frac{1}{2m}g^{\mu\nu}(p_{\mu}-qA_{\mu})(p_{\nu}
-qA_{\nu}) \nonumber \\
&=& \frac{H_1}{m} +\frac{1}{2m}\frac{\Delta}{\Sigma}p^{2}_{r}
+\frac{1}{2m}\frac{p^{2}_{\theta}}{\Sigma},
\end{eqnarray}
where  $p_{r}$ and $p_{\theta}$ are generalized momenta, and $H_1$
is a function of $r$ and $\theta$ [68]:
\begin{eqnarray}
H_1 &=&\frac{1}{2}g_{tt}[f_1(E+qA_t)+f_2(L-qA_{\phi})]^2
\nonumber \\
&& +\frac{1}{2}g_{\phi\phi}[f_2(E+qA_t)+f_3(L-qA_{\phi})]^2 \nonumber \\
&& -g_{t\phi}[f_1(E+qA_t)+f_2(L-qA_{\phi})] \nonumber \\
&& \cdot [f_2(E+qA_t)+f_3(L-qA_{\phi})].
\end{eqnarray}
Here, $f_1$, $f_2$ and $f_3$ are functions of $r$ and $\theta$ as
follows:
\begin{eqnarray}
f_1 &=& \frac{g_{\phi\phi}}{c^2(g_{tt}g_{\phi\phi}-g^{2}_{t\phi})}, \\
f_2 &=& \frac{g_{t\phi}}{c(g_{tt}g_{\phi\phi}-g^{2}_{t\phi})}, \\
f_3 &=& \frac{g_{tt}}{g_{tt}g_{\phi\phi}-g^{2}_{t\phi}}.
\end{eqnarray}
$E=-p_t$ is a constant energy of the particle, and $L=p_{\phi}$ is
a constant angular momentum of the particle. $p_t$ and $p_{\phi}$
are generalized momenta, which  satisfy the relations
\begin{eqnarray}
&& \dot{t} = \frac{\partial H}{\partial p_t}= -f_1(E+qA_t)-f_2(L-qA_{\phi}), \\
&& \dot{\phi} = \frac{\partial H}{\partial p_{\phi}}=
f_2(E+qA_t)+f_3(L-qA_{\phi}).
\end{eqnarray}
Because the 4-velocity
$U^{\alpha}=(c\dot{t},\dot{r},\dot{\theta},\dot{\phi})$ is always
identical to the constant $U^{\alpha}U_{\alpha}=-c^2$, the
Hamiltonian (5) remains invariant and obeys the constraint
\begin{equation}
H = -\frac{1}{2} mc^2.
\end{equation}
In fact, this third invariable quantity corresponds to the rest
mass of the particle.

For simplicity, $c$ and $G$ take geometrized units: $c=G=1$.
Dimensionless operations to the Hamiltonian (5) are carried out
thorough a series of scale transformations: $r\rightarrow rM$,
$t\rightarrow tM$, $\tau\rightarrow \tau M$, $a\rightarrow aM$,
$E\rightarrow Em$, $p_r\rightarrow mp_r$, $L\rightarrow mML$,
$p_{\theta}\rightarrow mMp_{\theta}$, $q\rightarrow mq$,
$B\rightarrow B/M$ and $H\rightarrow mH$. Note that no scale
transformation is given to the inductive charge $Q$. When these
treatments are employed, $M$ and $m$ in all the above-mentioned
expressions are eliminated or taken as geometrized units: $M=m=1$.
The horizon event of the black hole exists for $|a|\leq 1$.
For convenience, we take $Q^{*}=qQ$ and
$B^{*}=qB$.

\section{Effective potential and stable circular orbits}

Apart from the three integrals (10)-(12) in the dimensionless
Hamiltonian (5), the fourth constant related to the particles'
azimuthal motion is absent in general when the external magnetic
field forces are included. The absence of the fourth constant is
mainly caused by the $g_{\phi\phi}$ term in Eq. (4) rather than
the $g_{tt}$ term  in Eq. (3). Because $g_{tt}$ is only a function
of $r$, it does not destroy the presence of the fourth constant.
However, $g_{\phi\phi}$ is a function of $r$ and $\theta$ and
therefore the Hamilton-Jacobi equation of Eq. (5) has no separable
form of variables $r$ and $\theta$. This leads to the absence of
the fourth constant. Of course, the $g_{t\phi}$ terms being
functions of $r$ and $\theta$ in Eqs. (3) and (4) also have some
contributions to the absence of the fourth constant. In other
words, the main contribution to the absence of the fourth constant
in the system (5) comes from the external magnetic fields
associated with $B^{*}$. The inductive charges associated with
$Q^{*}$ also exert some influences on the absence of the fourth
constant. Thus, the dimensionless Hamiltonian (5) is
non-integrable. However, it can be integrable for some particular
cases. For instance, radial motions of charged particles on the
equatorial plane $\theta=\pi/2$ are integrable. The radial motions
are described in terms of effective potential $V$, i.e., the
expression of $E$ obtained from Eqs. (5) and (12) with
$p_r=p_{\theta}=0$:
\begin{equation}
V=E=\frac{B}{2 A}+\sqrt{\frac{B^{2}+4 A C+2 A}{4 A^{2}}},
\end{equation}
where $A$, $B$ and $C$ are expressed as
\begin{eqnarray}
A &=& -\frac{1}{2}(f_{1}^{2} g_{tt}+f_{2}^{2} g_{\phi\phi}-2 f_{1} f_{2} g_{t\phi}), \nonumber \\
B &=& \mathrm{B}_1+\mathrm{B}_2+B_3,  \nonumber \\
C &=& \frac{1}{2} g_{tt} C_1+\frac{1}{2} g_{\phi\phi} C_2-g_{t\phi} C_3, \nonumber \\
B_1 &=& g_{tt} q A_{t} f_{1}^{2}+g_{tt} f_{1} f_{2} L-g_{tt} f_{1} f_{2} q A_{\phi}, \nonumber \\
B_2 &=& g_{\phi\phi} q A_{t} f_{2}^{2}+g_{\phi\phi} f_{2} f_{3} L-g_{\phi\phi} f_{2} f_{3} q A_{\phi}, \nonumber \\
B_3 &=& -2 g_{t\phi} q A_{t} f_{1} f_{2}-g_{t\phi} f_{1} f_{3} L+g_{t\phi} q A_{\phi} f_{1} f_{3} \nonumber \\
&& -g_{t\phi} f_{2}^{2} L+g_{t\phi} q A_{\phi} f_{2}^{2}, \nonumber \\
C_1 &=& f_{1}^{2} q^{2} A_{t}^{2}+2 f_{1} f_{2} q L A_{t}-2 f_{1} f_{2} q^{2} A_{t} A_{\varphi}\nonumber \\
&& +f_{2}^{2}\left(A_{\phi}^{2} q^{2}-2 A_{\phi} L q+L^{2}\right), \nonumber \\
C_2 &=& f_{2}^{2} q^{2} A_{t}^{2}+2 f_{2} f_{3} q L A_{t}-2 f_{2} f_{3} q^{2} A_{t} A_{\phi}\nonumber\\
&&+f_{3}^{2}\left(A_{\phi}^{2} q^{2}-2 A_{\phi} L q+L^{2}\right), \nonumber \\
C_3 &=& A_{\phi}^{2} f_{2} f_{3} q^{2}-A_{\phi} A_{t} f_{1} f_{3} q^{2}-A_{\phi} A_{t} f_{2}^{2} q^{2}\nonumber\\
&&+A_{t}^{2} f_{1} f_{2} q^{2}-2 A_{\phi} f_{2} f_{3} L q+A_{t} f_{1} f_{3} L q\nonumber\\
&&+A_{t} f_{2}^{2} L q+f_{2} f_{3} L^{2}. \nonumber
\end{eqnarray}
The local minimal values of the effective potential correspond to
stable circular orbits, which satisfy the relation $dr/d\tau=0$
and the following conditions
\begin{equation}
\frac{d V}{d r}=0,
\end{equation}
\begin{equation}
\frac{d^{2} V}{d r^{2}} \geq 0.
\end{equation}
When the equality sign (=) is taken in Eq. (15), the innermost
stable circular orbit (ISCO) is present.

Taking parameters $L=2\sqrt{3}$, $a=0.1$, and $Q^{*}=2\times
10^{-4}$ (If $q=0.1$ and $B_W=0.01$, then $Q=0.002$ is the Wald
charge), we plot the effective potentials for several different
magnetic parameters $B^{*}$ in Fig. 1. When the magnetic parameter
$B^{*}$ increases, the left shape of the effective potential goes
away from the black hole, and the shape of the effective potential
is not altered. The energies of the unstable or stable circular
orbits become smaller. That is to say, the effective potential for
a larger value of $B^{*}$ is below that for a smaller value of
$B^{*}$. However, the radii of the stable circular orbits in Table
1 get larger as $B^{*}$ increases.

An increase of the inductive charge parameter $Q^{*}$ does not
alter the shape of the effective potential, but makes the left
shape of the effective potential go away from the black hole in
Fig. 2. Meantime, the energies of the unstable or stable circular
orbits decrease, but the radii of the stable circular orbits
increase in Table 2.

Fig. 3 clearly describes the dependence of the effective potential
on the black hole's spin $a$. The energies of the stable circular
orbits increase when $a$ gets larger. The radii of the stable
circular orbits always increase (see also Table 3).

How does the effective potential vary as the
particle's angular momentum $L$ increases? The effective potential
for a larger value of $L$ is always over that for a smaller value
of $L$, as shown in the Kerr spacetime of Figs. 4 (b) and (c).
Note that there are critical values of $L$ corresponding to the
ISCOs colored Red in Table 3, such as $L=3.4641$ for the
Schwarzschild spacetime with $a=0$. When the angular momenta $L$
are larger than the critical values, the stable circular orbits
are present in Table 3. However, no stable circular orbits exist
for $L$ less than the critical values. As $a$ or $L$ increases,
the radii of the stable circular orbits also increase.

Although the radii of the stable circular orbits increase with an
increase of $a$, the radii of the ISCOs become smaller in Table 3.
In addition, the radii of the ISCOs depend on the sign of the
particle's angular momentum  as well as the magnitude of the
particle's angular momentum. When $L>0$ (for this case, the spin
direction of the black hole is consistent with the particle's
angular momentum), the considered orbits are called as direct
orbits.  When $L<0$ (for this case, the spin direction of the
black hole is opposite to the particle's angular momentum), the
considered orbits are called as retrograde orbits [38]. Given
parameters $a$, $Q^{*}$ and $B^{*}$, the radii of the ISCOs for
the retrograde orbits are larger than those for the direct orbits.
Any one of the parameters $a$, $Q^{*}$ and $B^{*}$ increases, the
radii of the ISCOs for the retrograde orbits or the direct orbits
decrease. More details on the ISCOs are listed in Tables 4-6.

\section{Numerical investigations}

Without loss of generality, the Hamiltonian system (5) for the
description of the motion of charged particles at the
non-equatorial plane is nonintegrable and has no analytical
solutions. Numerical integration schemes are convenient to solve
this system. Particularly for obtaining the numerical solutions of
the Hamiltonian problem, symplectic integrators are naturally a
prior choice. Because explicit symplectic integrators are
generally superior to implicit ones at same order in computational
efficiency, their applications also remain a high priority. Owing
to the difficulty in the separation of variables or the separation
of two integrable parts in curved spacetimes, the implicit
symplectic integrators rather than the explicit ones are suitably
applicable to the curved spacetimes in general [41, 54-58].
Recently, Wang et al. [64-66] split the Hamiltonians of
non-rotating black holes surrounded by external magnetic fields
into several parts with analytical solutions as explicit functions
of proper time $\tau$, and successfully constructed the explicit
symplectic integrators for these non-rotating black holes. More
recently, the authors of [67] gave a time transformation to the
Kerr geometry, and designed the explicit symplectic integrators
for the time-transformed Hamiltonian with a desired splitting
form. The time-transformed explicit symplectic integrators were
applied to study the dynamics of charged particles moving around
the Kerr black hole surrounded by external magnetic fields without
the inductive charge $Q$ [68]. Following the two works [67, 68],
we use the explicit symplectic integrators to the Hamiltonian
problem (5).

\subsection{Explicit symplectic integrators}

The authors of [67] introduced a time transformation function
\begin{equation}
d\tau=g(r,\theta)dw, ~~~~   g(r,\theta)=\frac{\Sigma}{r^2},
\end{equation}
where $w$ is a new coordinate time unlike the original coordinate
time $t$. The Hamiltonian (15) becomes
\begin{equation}
K=g(H+p_0)=\frac{\Sigma}{r^2} (H_1+p_0)
+\frac{\Delta}{2r^2}p^{2}_{r} +\frac{1}{2r^2}p^{2}_{\theta}.
\end{equation}
The new Hamiltonian $K$ is a time-transformed Hamiltonian, where
the proper time $\tau$ is viewed as a coordinate $q_0=\tau$ and
its corresponding momentum  is $p_0=-H=1/2\neq p_t$. In this case,
$K$ is always identical to zero for any coordinate time $w$, i.e.,
\begin{equation}
K\equiv 0.
\end{equation}
Now, the time-transformed Hamiltonian $K$ in Eq. (17) is split
into five parts
\begin{equation}
K=K_1+K_2+K_3+K_4+K_5,
\end{equation}
where all sub-Hamiltonians are expressed as
\begin{eqnarray}
K_1 &=& \frac{\Sigma}{r^2} (H_1+p_0), \\
K_{2} &=& \frac{1}{2}p^{2}_{r},\\
K_{3} &=& -\frac{1}{r}p^{2}_{r},\\
K_{4} &=&
\frac{a^2}{2r^2}p^{2}_{r}, \\
K_{5} &=& \frac{1}{2r^2}p^{2}_{\theta}.
\end{eqnarray}
$K_2$, $K_3$ and $K_5$ are consistent with those of [67], but
$K_1$ and $K_4$ are not.

Each of the five sub-Hamiltonians $K_1$, $K_2$, $K_3$, $K_4$ and
$K_5$ is solved analytically, and its solutions are explicit
functions of the new coordinate time $w$. Operators associated to
the solutions of $K_1$, $K_2$, $K_3$, $K_4$ and $K_5$ are
$\hat{K}_1$, $\hat{K}_2$, $\hat{K}_3$, $\hat{K}_4$ and
$\hat{K}_5$, respectively. The solutions of the system (5)
advancing a new coordinate time step $\Delta w=h$ are given in
terms of an explicit second-order symplectic integrator
\begin{eqnarray}
S^{K}_2(h) &=& \hat{K}_5 (\frac{h}{2}) \circ
\hat{K}_4(\frac{h}{2})\circ \hat{K}_3(\frac{h}{2})\circ
\hat{K}_2(\frac{h}{2})\circ \hat{K}_1(h) \nonumber
\\ && \circ
\hat{K}_2(\frac{h}{2}) \circ \hat{K}_3(\frac{h}{2})\circ
\hat{K}_4(\frac{h}{2})\circ \hat{K}_5(\frac{h}{2}),
\end{eqnarray}
as was proposed in [67]. The second-order method easily yields a
fourth-order symplectic integrator [72]
\begin{equation}
S^{K}_4(h)=S^{K}_2(\gamma h)\circ S^{K}_2(\delta h)\circ
S^{K}_2(\gamma h),
\end{equation}
where $\delta=1-2\gamma$ and $\gamma=1/(2-\sqrt[3]{2})$.

In fact, the explicit symplectic algorithms (25) and (26) are
attributed to  the development of the time-transformed symplectic
method of [73] in the Kerr spacetime and its extension. The method
of Mikkola [73] aims to exhibit good performance of symplectic
integrators for close encounters of objects or high orbital
eccentricities in the solar system. In the idea of Mikkola, these
integrators use fixed time steps and remain symplectic for the new
time, but adaptive time steps for the original time. However, the
time steps in the method of [67] including the present integrators
(25) and (26) are approximately invariant for the proper time
$\tau$ because $g\approx 1$ and $\Delta\tau\approx g\Delta
w\approx\Delta w=h$ for $r\gg2$ in Eq. (16). As the authors of
[67] claimed, the time transformation mainly aims to eliminate the
function $\Sigma$ in the denominators of the terms $p_r$ and
$p_{\theta}$ in the Hamiltonian $H$ and to cause the
time-transformed Hamiltonian $K$ to have the desired separable
form.

In comparison with S4, a fourth-order method implicit symplectic
algorithm (IM4) consisting of three second-order implicit midpoint
methods [56] is applied to the time-transformed Hamiltonian $K$.
The conventional fourth-order Runge-Kutta explicit integration
method (RK4) is also employed. Of course, IM4 and RK4 are suitable
for the original Hamiltonian (5).

The new coordinate time step is given by $h=1$. The parameters are
$E=0.9935$, $L=4.6$, $a=0.5$, $B^{*}=1\times10^{-3}$, and $Q^{*}=1\times10^{-3}$.
The initial conditions are $\theta=\pi/2$ and $p_{r}=0$. If the
initial separation $r$ is given, then the initial value
$p_{\theta}>0$ is obtained from Eq. (12). We take $r=55$ for Orbit
1, and $r=75$ for Orbit 2. When the three algorithms S4, IM4 and
RK4 independently integrate the two orbits in the system (17), the
evolutions of $K$ in Eq. (18) with integration time $w$ are shown
in Figs. 5 (a) and (b). The explicit symplectic method S4 and the
implicit symplectic algorithm IM4 do not show secular drifts in
Hamiltonian errors, but RK4 does. In addition, S4 and IM4 are
almost the same, and have two orders of magnitude smaller errors
than RK4. Accuracy of each algorithm for Orbit 2 in Fig. 5(b) has
an advantage over that for Orbit 1 in Fig. 5(a). Is this result
because Orbit 2 is regular and Orbit 1 is chaotic?  In fact, Orbit
1 is a regular Kolmogorov-Arnold-Moser (KAM) torus, but Orbit 2
exhibits the chaoticity, as described through Poincar\'{e} section
at the plane $\theta=\pi/2$ with $p_{\theta}>0$ in Fig. 5(c). The
result on the preference of accuracy of each algorithm for Orbit 2
over that for Orbit 1 is because Orbit 1 has a larger average
period than Orbit 2. Although both orbits are not exactly periodic
and Orbit 2 are chaotic, they have approximately average periods.
Based on good computational efficiency, S4 is employed in the
later studies.

\subsection{Dynamics of generic orbits}

Let us consider the effect of a small change of the inductive
charge parameter $Q^{*}$ on the orbital dynamics. If $Q^{*}=1\times10^{-3}$
in Fig. 5(c) gives place to $Q^{*}=0$, no chaos exists in Fig.
6(a). When $Q^{*}=5\times10^{-4}$, all the orbits in Fig. 6(b) are still
regular. As the inductive charge parameter increases to
$Q^{*}=6\times10^{-4}$ in Fig. 6(c), the pink orbit with the initial
separation $r=100$ is chaotic. For $Q^{*}=8\times10^{-4}$ in Fig. 6(d),
Orbit 2 and the pink orbit with the initial separation $r=100$ are
chaotic. As the inductive charge parameter increases and takes the
Wald charge $Q^{*}=2aBq=1\times10^{-3}$ in Fig. 5(c), chaos becomes
stronger from the global phase-space structure. These facts show
that a small increase of the inductive charge parameter $Q^{*}$
can easily induce chaos. An explanation to this result is given
here. The inductive charges in the vicinity of the Wald charge are
so small that they do not contribute to the spacetime curvature,
but can exert somewhat important influences on the motions of
charged particles and even enhance the chaotic properties. The
inductive charges have small contributions to the absence of the
fourth constant although the external magnetic fields have main
contributions, as is claimed above. When $B^{*}$ is given an
appropriate value and the ratio of the particle's charge $q$ to
the particle's mass $m$ (i.e., $q/m$) is large enough, the
inductive charges are possible to bring a contribution to the
occurrence of chaos.

A minor change of the magnetic parameter $B^{*}$ has an important
effect on the orbital dynamics. With charge $B^{*}$ increasing,
the evolution of orbits transits from regular KAM tori for
$B^{*}=3\times10^{-4}$ in Fig. 7(a) to chaos for $B^{*}=8\times10^{-4}$ in Fig.
7(b), and to stronger chaos for $B^{*}=1.1\times10^{-3}$ in Fig. 7(c). An
increase of $B^{*}$ means that of the Lorentz force and therefore
enhances the strength of chaos.

The above demonstrations mainly focus on how the two parameters
$Q^{*}$ and $B^{*}$ exert influences on the dynamical behavior of
orbits. What about the effect of the black hole spin $a$ on the
orbital dynamics? Fig. 8 gives an answer to this question. It is
found that the chaotic properties are gradually weakened and ruled
out as the dragging effects of the spacetime by the rotating black
hole increase. This fact supports the result of [12]. It is shown
again that an increase of the inductive charge parameter $Q^{*}$
is helpful to induce chaos for a given value $a$. Similarly, an
increase of the particle's angular momentum $L$ also results in
weakening and suppressing the chaotic properties, as shown in Fig.
9.

As is well known, chaos is stronger when the larger the particle's
energy $E$ increases. This result is confirmed by fast Lyapunov
indicators (FLIs) in Fig. 10. Here, computations of the FLIs are
based on the method of [74]. The FLI is the logarithm of the ratio
of the separation between two nearby trajectories $d(\tau)$ at
proper time $\tau$ to the starting separation $d(0)$:
\begin{equation}\label{Eq:fli}
FLI=\log_{10}\frac{d(\tau)}{d(0)}.
\end{equation}
Different growths of separation $d(\tau)$ with proper time $\tau$
allow one to distinguish between ordered and chaotic orbits. A
slowly polynomial or  algebraical increase of the separation
indicates the regularity of the considered bounded orbit for
$E=0.9925$ in Fig. 10. However, a rapidly exponential increase of
the separation turns out to be the characteristic of chaoticness
of the considered bounded orbit for $E=0.9935$. The FLI for
$E=0.995$ is less than for $E=0.997$ after the integration time
$w=10^{6}$ or $\tau=10^{6}$, therefore, the former chaoticity is
weaker than the latter one. That is, an increase of the energy $E$
gives rise to enhancing strength of chaos.

We find that the FLIs are always smaller than 3.5 for the regular
case, whereas larger than this value for the chaotic case when the
integration time reaches $w=10^{6}$. Now, we employ the technique
of FLIs to trace how a small variation of one parameter affects a
dynamical transition from order to chaos. Only one of the
parameters is given many different values, and the initial
conditions (except $p_{\theta}$) and the other parameters are
fixed. Each FLI is obtained after the integration time $w=10^{6}$.
The transition from order to chaos occurs when $Q^{*}\geq0.00056$ (Fig.
11(a)), $B^{*}\geq 0.000844$ (Fig. 11(b)), or $E\geq0.99379$ (Fig.
11(c)). However, the transition from chaos to order occurs when
$L\geq5.84789$ (Fig. 11(d)). That is, the strength of chaos is
enhanced as one of the parameters $Q^{*}$, $B^{*}$ and $E$
increases, but weakened as the parameter $L$ increases. The
effects of variations of these parameters on the orbital dynamics
described by the technique of FLIs are consistent with those
described by the technique of Poincar\'{e} sections.

The transition from order to chaos occurs when the black hole's
spin $a\geq0.046$ in Fig. 11(e). Namely, an increase of $a$ leads
to strong chaos. The result is consistent with that of [68], but
unlike that of Fig. 8 in which the dragging effects of the
spacetime weaken the chaotic properties from the global
phase-space structures. The different results between Figs. (8)
and 11(e) are due to distinct choices of the initial conditions
and other parameters. Perhaps, the dependence of the dynamical
behavior on the spin may be different if  the chosen initial
conditions and other parameters are varied.

\section{Conclusions}

In this paper, we mainly focus on  studying the dynamics of
charged particles around the Kerr black hole immersed in an
external electromagnetic field, which can be considered as a tidal
environment.

At first, we discuss the radial motions of charged particles on
the equatorial plane through the effective potential. We trace how
the dynamical parameters exert influences on the effective
potential. It is found that the particle energies at the local
maxima values of the effective potentials increase with an
increase of the black hole spin and the particle angular momentum,
whereas decrease as one of the inductive charge parameter and
magnetic field parameter increases. In addition, the radii of
stable circular orbits on the equatorial plane always increase.
However, the radii of ISCOs are decreasing as any
one of the black hole spin $|a|$, inductive charge parameter
$Q^{*}$ and uniform magnetic field  parameter $B^{*}$ is
increasing.

Then, we investigate the motions of charged particles on the
non-equatorial plane using a time-transformed explicit symplectic
integrator. The effects of small variations of the parameters on
the orbital regular and chaotic dynamics are studied through the
techniques of Poincar\'{e} sections and fast Lyapunov indicators.
As a result, the dynamics is  sensitive dependence on a small
variation of any one of the inductive charge parameter, magnetic
field  parameter, energy and angular momentum. Chaos is easily
induced as these dynamical parameters increase. When the dragging
effects of the spacetime increase, the chaotic properties are not
always enhanced or weakened under some circumstances.

The theoretical work may have potential astrophysical
applications. The unstable or stable circular orbits and ISCOs
would be helpful to study some accretion disks. The theory of
chaotic scattering in the combined effective potential and the
asymptotically uniform magnetic field would be applicable to
explaining the mechanism hidden behind the charged particle
ejection. The existence of the magnetic fields involving the
induced electric field may be demonstrated through observational
evidences.

\section*{Acknowledgments}

The authors are very grateful to three referees for valuable
comments and useful suggestions. This research has been supported
by the National Natural Science Foundation of China [Grant Nos.
11973020 (C0035736) and 12133003], the Special Funding for Guangxi
Distinguished Professors (2017AD22006), and the National Natural
Science Foundation of Guangxi (Nos. 2018GXNSFGA281007, and
2019JJD110006).

\newpage

\begin{table*}
\begin{center}
\small \caption{Radii $R$ of stable circular orbits  in Fig. 1.}
\begin{tabular}{ccccc}\hline
Parameter                    & $B^\ast=0.001$    & $B^\ast=0.002$ & $B^\ast=0.003$   & $B^\ast=0.004$  \\
\hline
 $R$     & 7.4595              & 7.5165         & 7.5693           & 7.6177           \\
\hline
Parameter                    & $B^\ast=0.005$    & $B^\ast=0.006$ & $B^\ast=0.007$   & $B^\ast=0.008$  \\
\hline
 $R$     & 7.6618              & 7.7013        & 7.7365           & 7.7673                     \\
\hline
\end{tabular} 
\end{center}
\end{table*}

\begin{table*}
\begin{center}
\small \caption{Radii $R$ of stable circular orbits in Fig. 2.}
\begin{tabular}{ccccc}\hline
 $a=0.1$                    & $B^\ast=0.001$   & $B^\ast=0.002$  \\
\hline
$Q^\ast=0$                  & 7.4559                    & 7.5129                   \\
\hline
$Q^\ast=0.0001$     & 7.4577                    & 7.5147                   \\
\hline
$Q^\ast=0.0004$     & 7.4631                    & 7.5201                   \\
\hline
\end{tabular} 
\end{center}
\end{table*}

\begin{table*}
\begin{center}
\caption{Radii $R$ of stable circular orbits  in the Kerr
spacetime of Fig. 4. The radii $R$ colored Red are those of the
innermost stable circular orbits (ISCOs).}
\begin{tabular}{cccccc}\hline
$a$   & $L$         & $R$             \\
\hline
    & \textcolor{red}{3.4641}    & \textcolor{red}{6}    \\

0   & 4.0       & 12       \\

    & 4.5       & 16.5876  \\
\hline
& \textcolor{red}{3.2640}   & \textcolor{red}{5.3294}  \\
0.2  & 4.0      & 12.5329 \\
 & 4.5      & 16.9761 \\
\hline
    & \textcolor{red}{3.0340}   & \textcolor{red}{4.6143} \\
0.4 & 4.0       & 12.9954   \\

    & 4.5       & 17.3336  \\
\hline
& \textcolor{red}{2.7559}   & \textcolor{red}{3.8290}  \\
0.6  & 2.8      & 4.5721  \\
  & 2.9      & 5.4096  \\
\hline
\end{tabular} 
\end{center}
\end{table*}

\begin{table*}
\begin{center}
\caption{Radii $R$ of the ISCOs. The parameters are $a=0.2$,
$Q^\ast=0.008$. DO represents a direct orbit, and RO denotes a
retrograde orbit. Note that $L$ with $E$ is not arbitrarily given
for the case of ISCOs, but it is for the case of stable circular
orbits. $R$, $E$ and $L$ are determined together in terms of the
conditions of ISCOs.  }
\begin{tabular}{ccccccccccc}\hline
          & $B^\ast=0.02$   & $B^\ast=0.02$ & $B^\ast=0.04$  &  $B^\ast=0.04$ & $B^\ast=0.06$  & $B^\ast=0.06$   \\
          & DO              & RO            & DO             & RO             & DO             & RO    \\
\hline
$R$       & 5.2406          & 6.4737        & 5.0411         & 6.1176         & 4.7844        & 5.8262  \\
\hline
$L$       & 3.1794          & -3.8201       & 3.1433          &-4.0808        & 3.1311        &-4.3879   \\
\hline
$E$       & 0.9078          & 0.9879        & 0.8885         & 1.0406         & 0.8726         & 1.10173    \\
\hline
\end{tabular} 
\end{center}
\end{table*}

\begin{table*}
\begin{center}
\caption{Similar to Table 4, but the parameter are $a=0.2$ and
$B^\ast=0.01$. }
\begin{tabular}{ccccccccc}\hline
          & $Q^\ast=0$    &  $Q^\ast=0$   &  $Q^\ast=0.001$  &  $Q^\ast=0.001$  & $Q^\ast=0.002$  & $Q^\ast=0.002$   \\
         & DO             & RO            & DO              & RO              & DO             & RO   \\
\hline
$R$       & 5.3276       & 6.599          & 5.3275          & 6.6193          & 5.3274         & 6.6043    \\
\hline
$L$       & 3.2224       & -3.7282        & 3.2209          &-3.7265          &3.2194          &-3.7247     \\
\hline
$E$       & 0.9226       & 0.9688         & 0.9222          & 0.9683          & 0.9217         & 0.9679      \\
\hline
\end{tabular} 
\end{center}
\end{table*}

\begin{table*}
\begin{center}
\caption{Similar to Table 4, but the parameter are $a=0.2$ and
$Q^\ast=0$. }
\begin{tabular}{ccccccccc}\hline
          & $B^\ast=0$    &  $B^\ast=0$   &  $B^\ast=0.02$  &  $B^\ast=0.02$  & $B^\ast=0.04$  & $B^\ast=0.04$   \\
         & DO              & RO           & DO              & RO              & DO             & RO   \\
\hline
$R$       & 5.3641         & 6.6713       & 5.2534          & 6.4653          & 5.0411         & 6.1218    \\
\hline
$L$       & 3.2641         & -3.6434      & 3.1912          &-3.8343          & 3.1546         &-4.0954     \\
\hline
$E$       & 0.9354         & 0.9485       & 0.9112          & 0.9917          & 0.8918         & 1.0445      \\
\hline
\end{tabular} 
\end{center}
\end{table*}

\newpage

\begin{figure*}[!ht]
\centering{
\includegraphics[width=15pc]{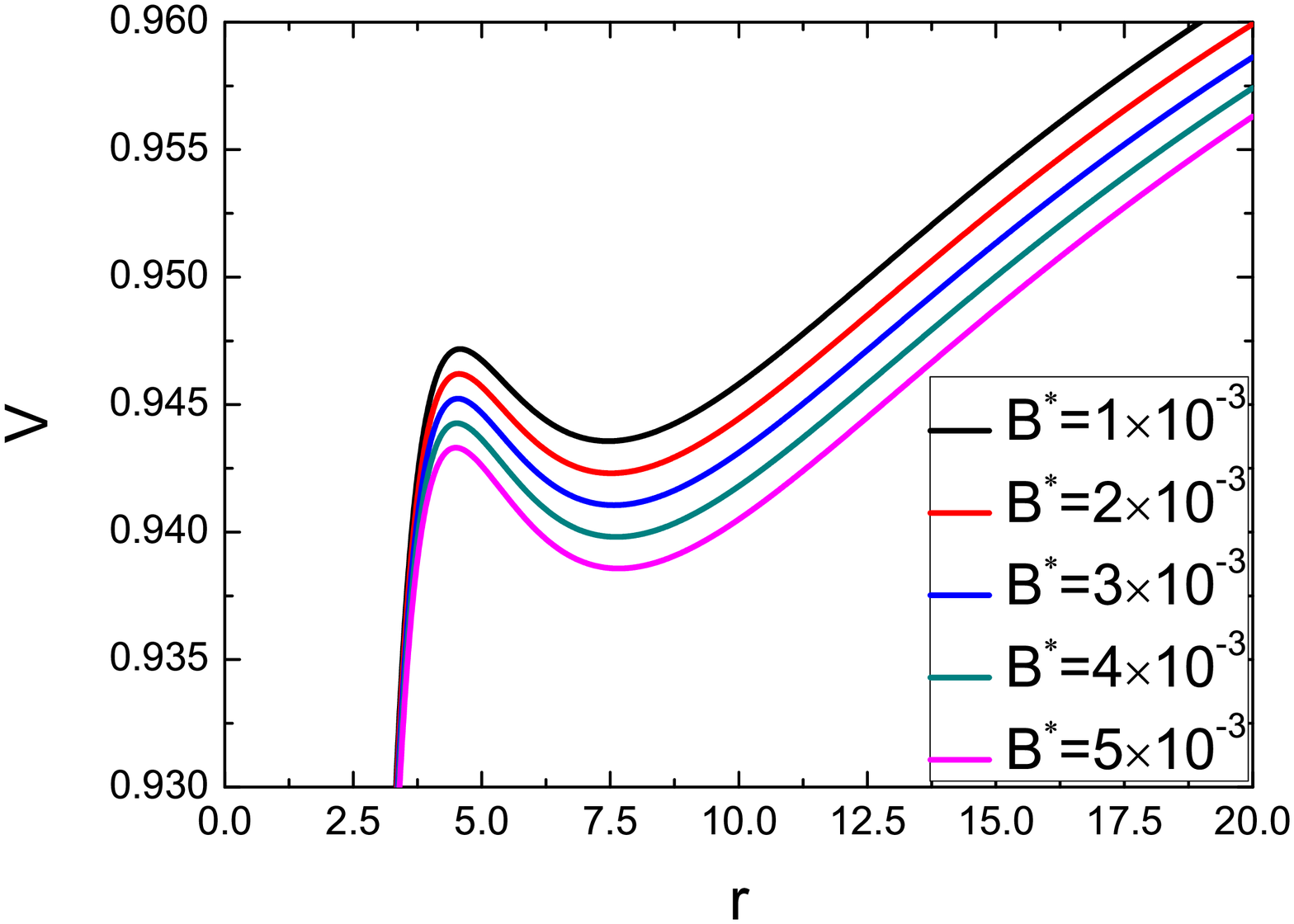}
\caption{Effective potentials for different uniform magnetic field
parameters $B^{*}$. The other parameters are $L=2\sqrt{3}$,
$a=0.1$ and $Q^\ast=0.0002$. }\label{figure1}}
\end{figure*}

\begin{figure*}[!ht]
\centering{
\includegraphics[width=15pc]{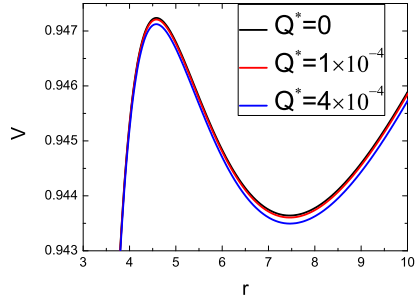}
\caption{Effective potentials for several inductive charge
parameters $Q^{*}$. The other parameters are $B^{*}=0.001$,
$a=0.1$ and $L=2\sqrt{3}$. }\label{figure2}}
\end{figure*}

\begin{figure*}[!ht]
\center{
\includegraphics[width=15pc]{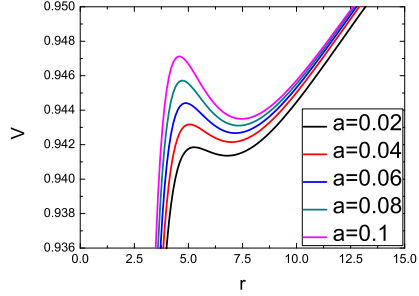}
\caption{Effective potentials for several black hole spins $a$.
The other parameters are $L=2\sqrt{3}$, $B^\ast=0.001$ and
$Q^\ast=0.004$. }\label{figure3}}
\end{figure*}

\begin{figure*}[!ht]
\centering{
\includegraphics[width=12pc]{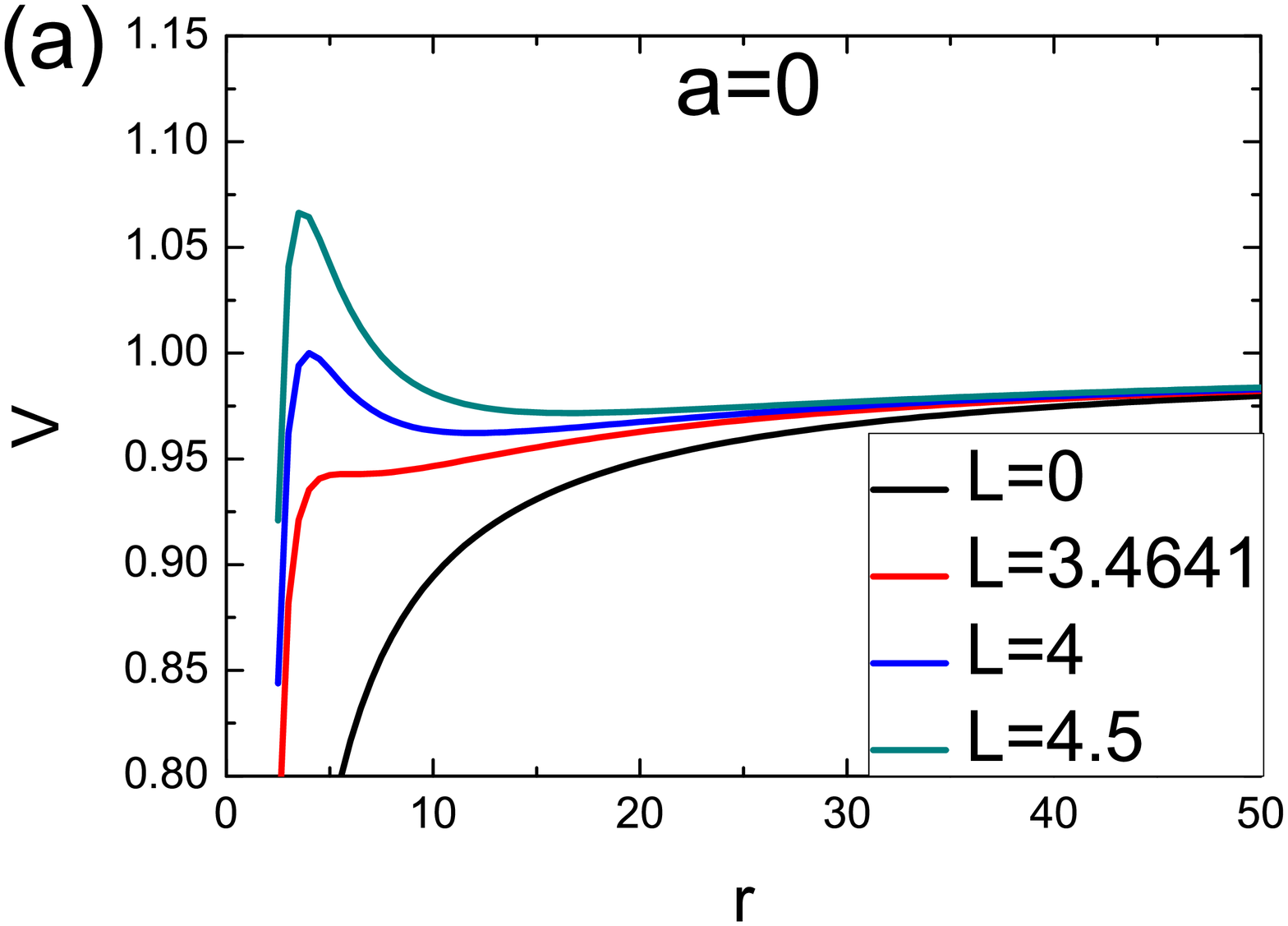}
\includegraphics[width=12pc]{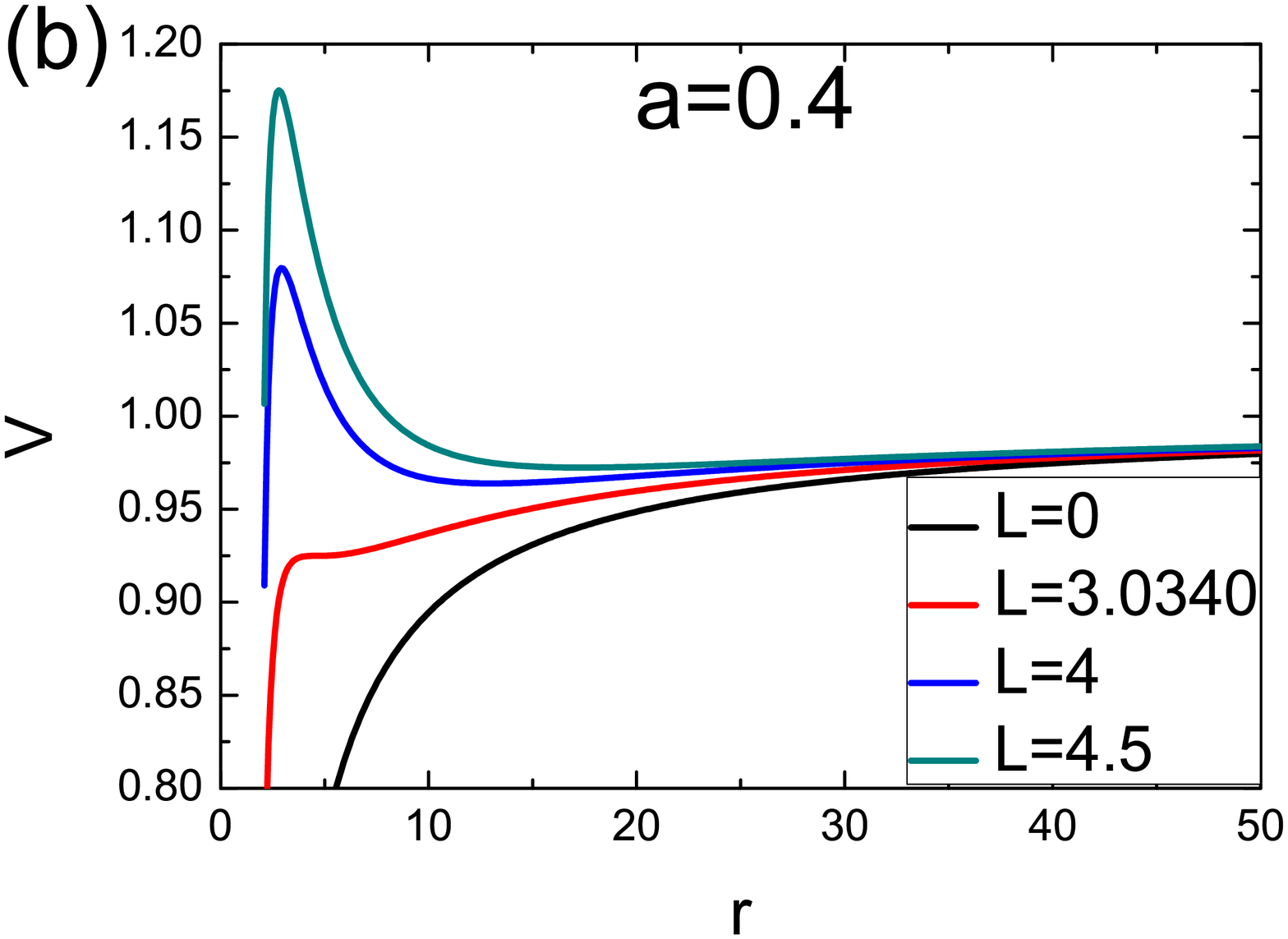}
\includegraphics[width=12pc]{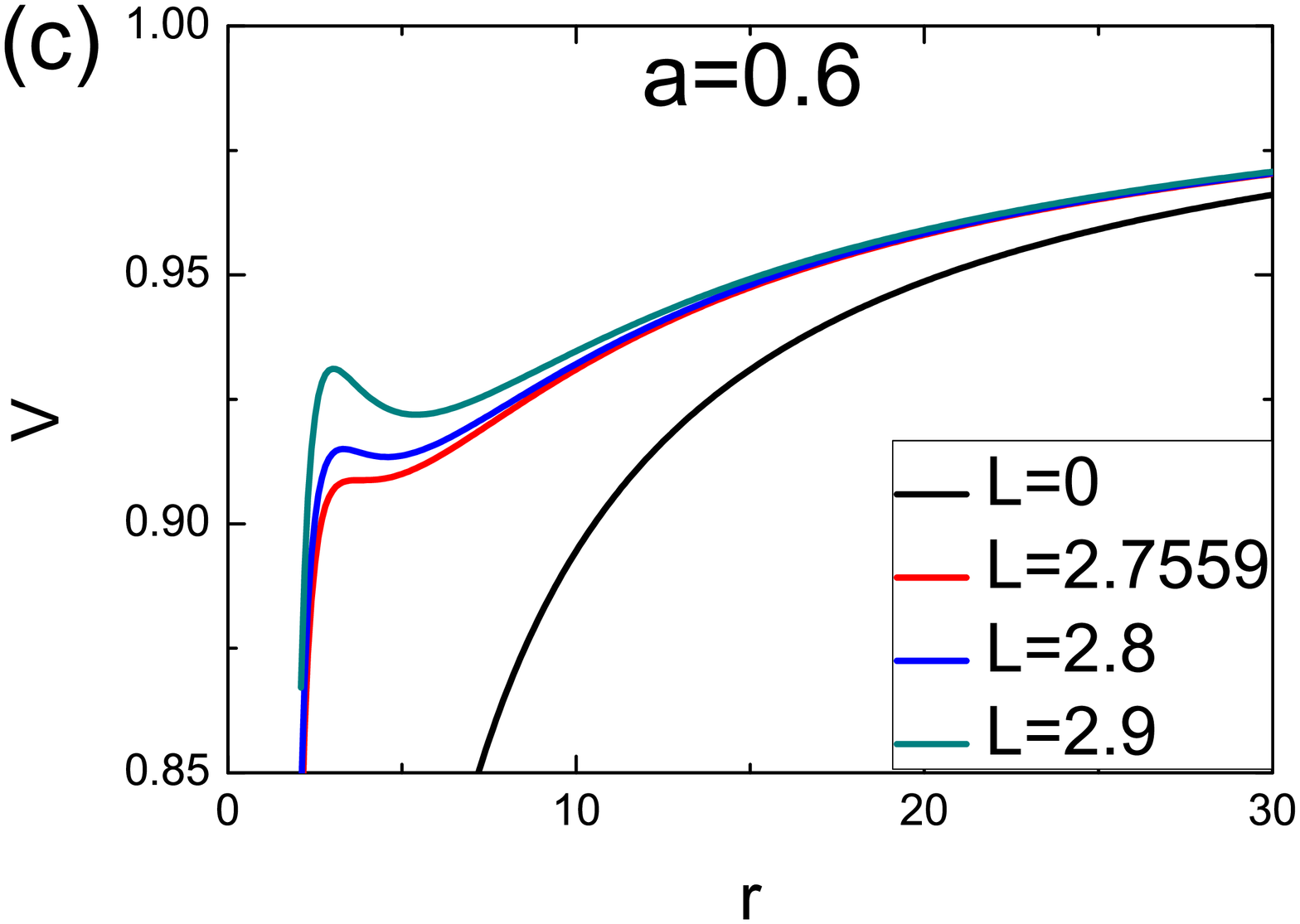}
\caption{Effective potentials for different angular momenta $L$ of
test particles. The other parameters are  $Q^\ast=0$ and
$B^\ast=0$.}\label{figure4}}
\end{figure*}

\begin{figure*}[!ht]
\centering{
\includegraphics[width=12pc]{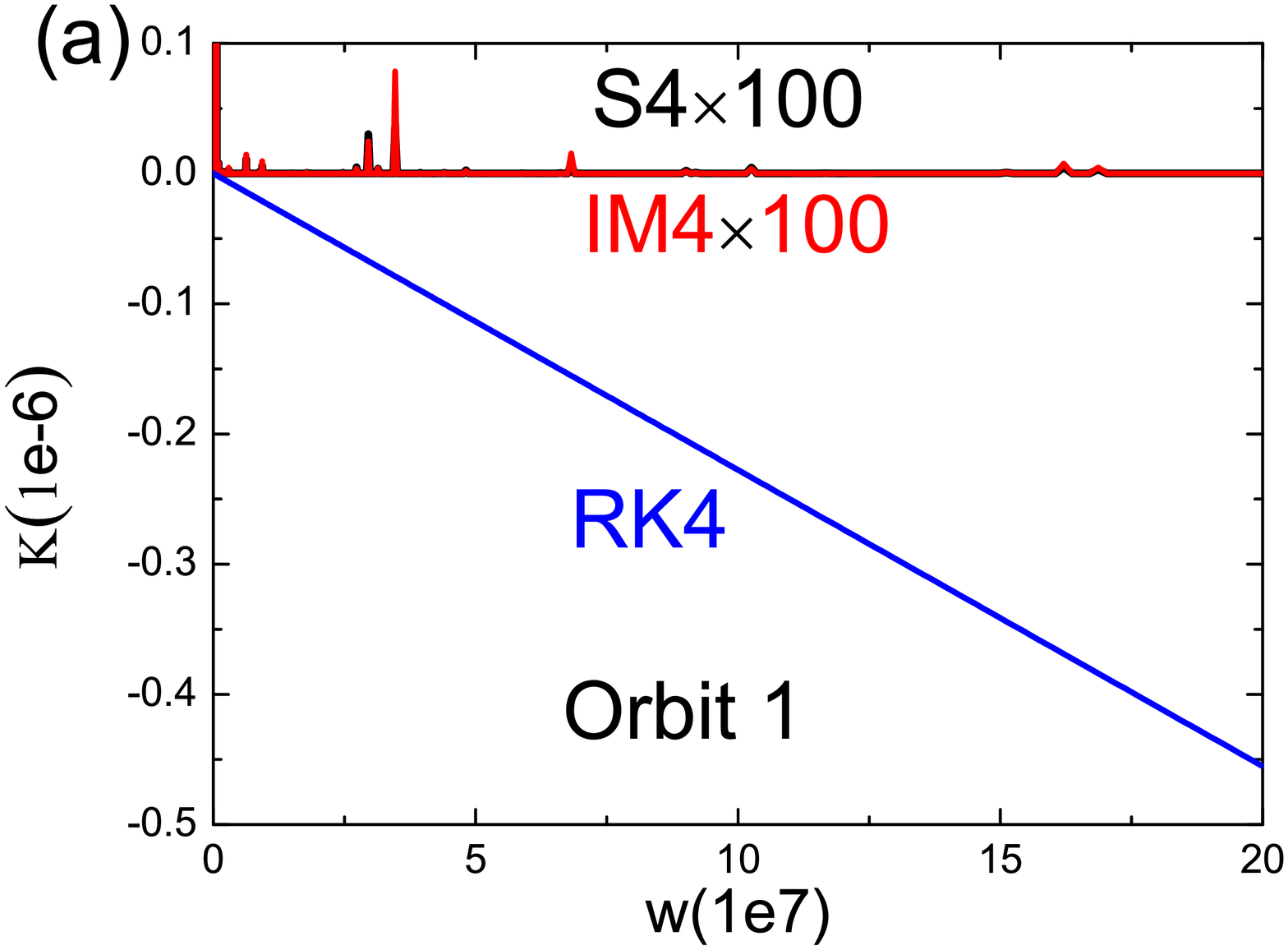}
\includegraphics[width=12pc]{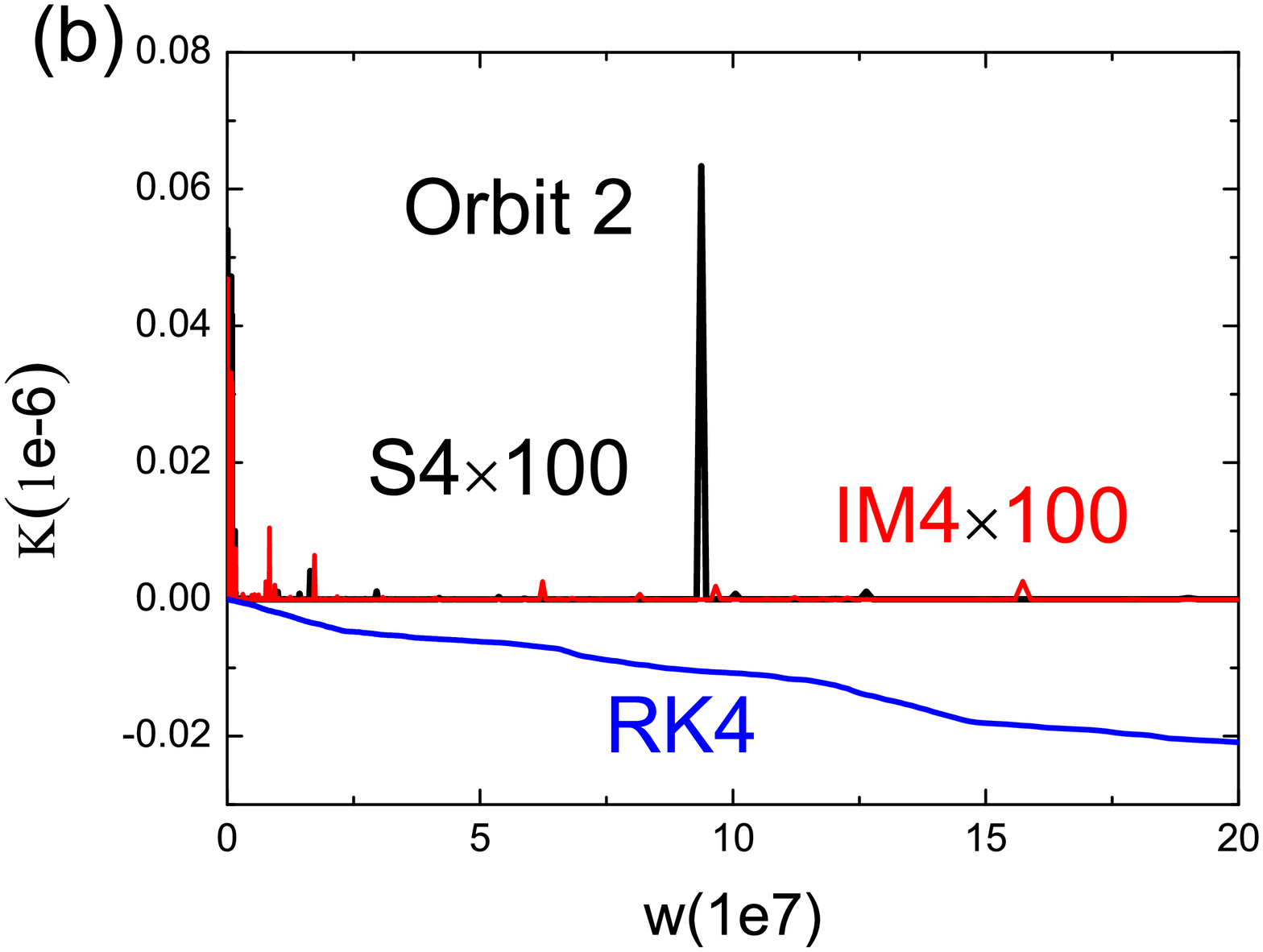}
\includegraphics[width=12pc]{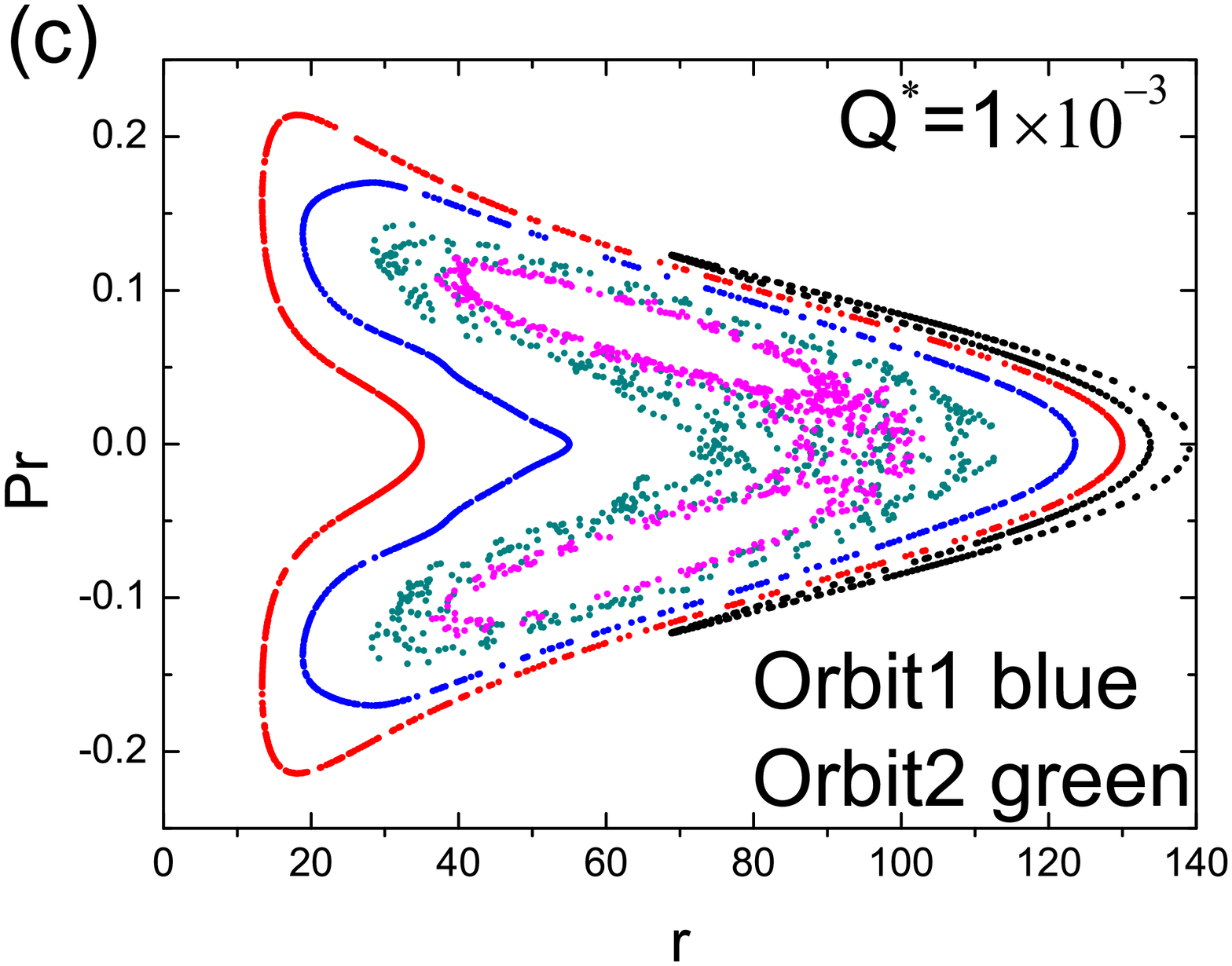}
\caption{(a) and (b): Accuracies of the time-transformed
Hamiltonian $K$ yielded by three methods S4, IM4 and RK4 acting on
Orbit 1 and Orbit 2. The time-step is $h=1$. Orbits 1 and 2 have
their initial separations $r=55$ and $r=75$, respectively. The
other initial conditions are $\theta=\pi/2$, $p_{r}$=0 and
$p_{\theta}>0$. The parameters are $E=0.9935$, $L=4.6$,
$B^\ast=0.001$,  $a=0.5$ and $Q^\ast=0.001$. Symbol ``$\times
100$" means that the plotted errors are enlarged 100 time compared
with the practical errors. (c) Poincar\'{e} sections at plane
$\theta=\pi/2$ with $p_{\theta}>0$. The black orbit has its
initial separation $r=25$, the red orbit has the initial
separation $r=35$, and the pink orbit has the initial separation
$r=100$.  Orbit 1 is regular, whereas Orbit 2 and the pink orbit
with the initial separation $r=100$ are chaotic. }\label{figure5}}
\end{figure*}

\begin{figure*}[!ht]
\centering{
\includegraphics[width=15pc]{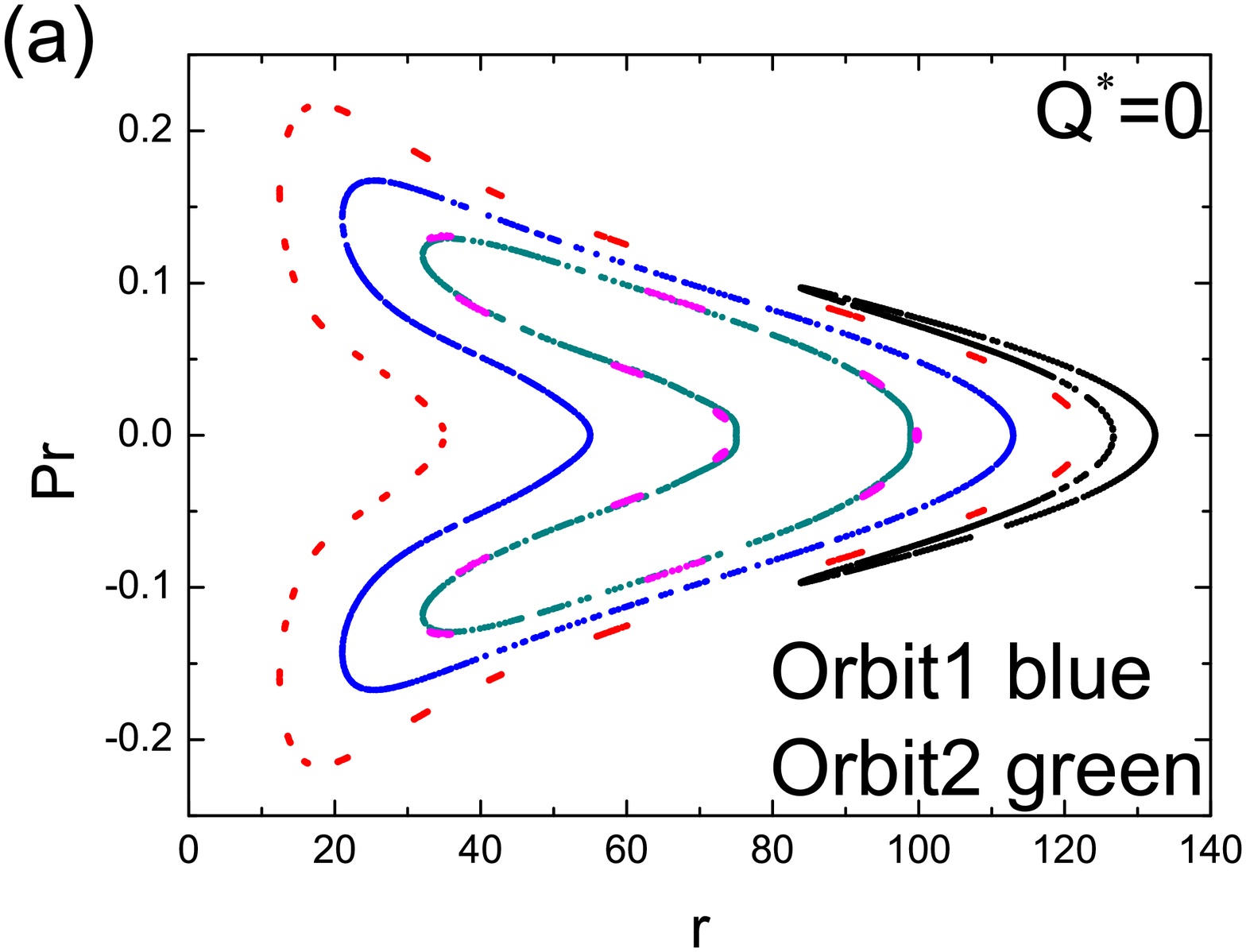}
\includegraphics[width=15pc]{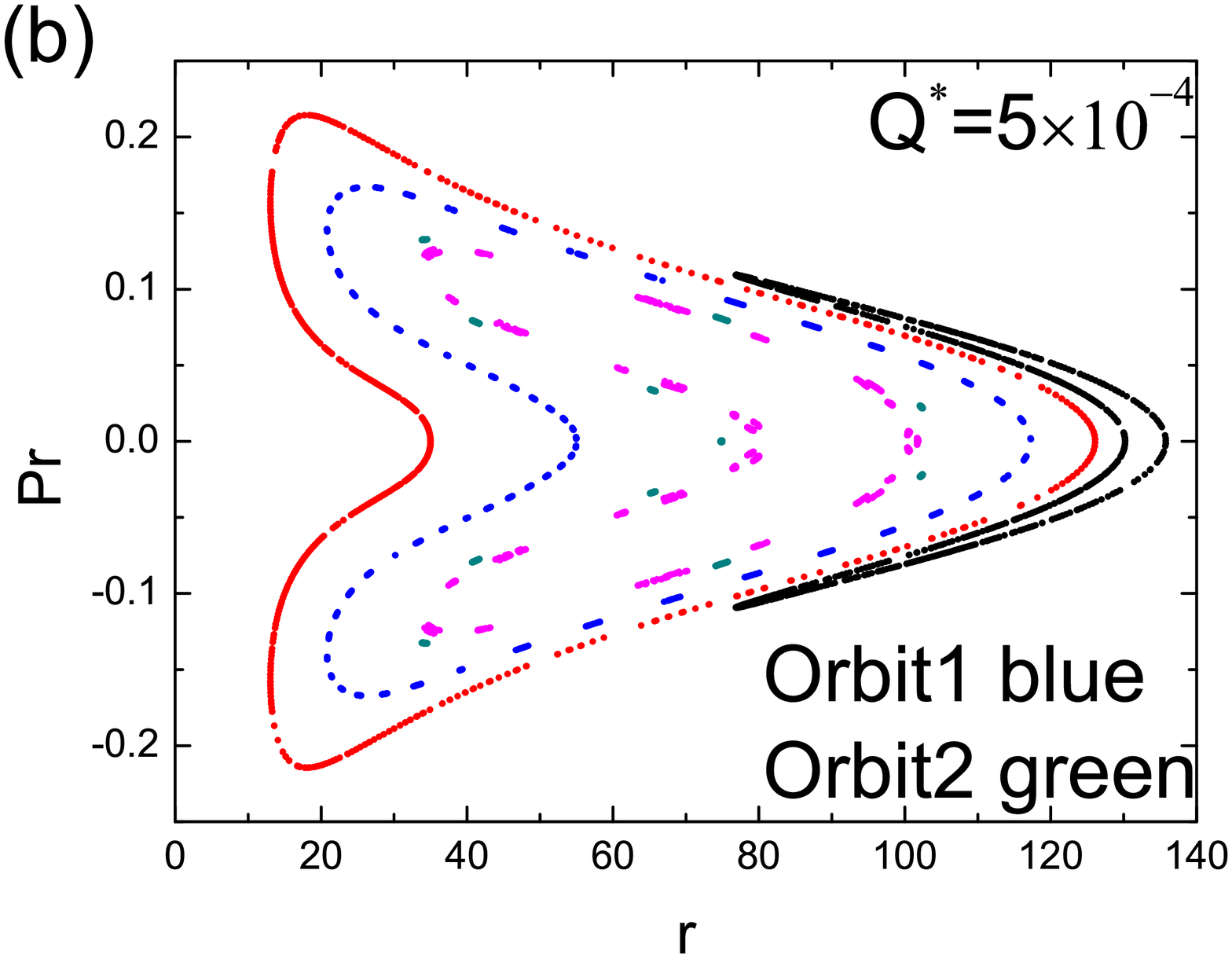}
\includegraphics[width=15pc]{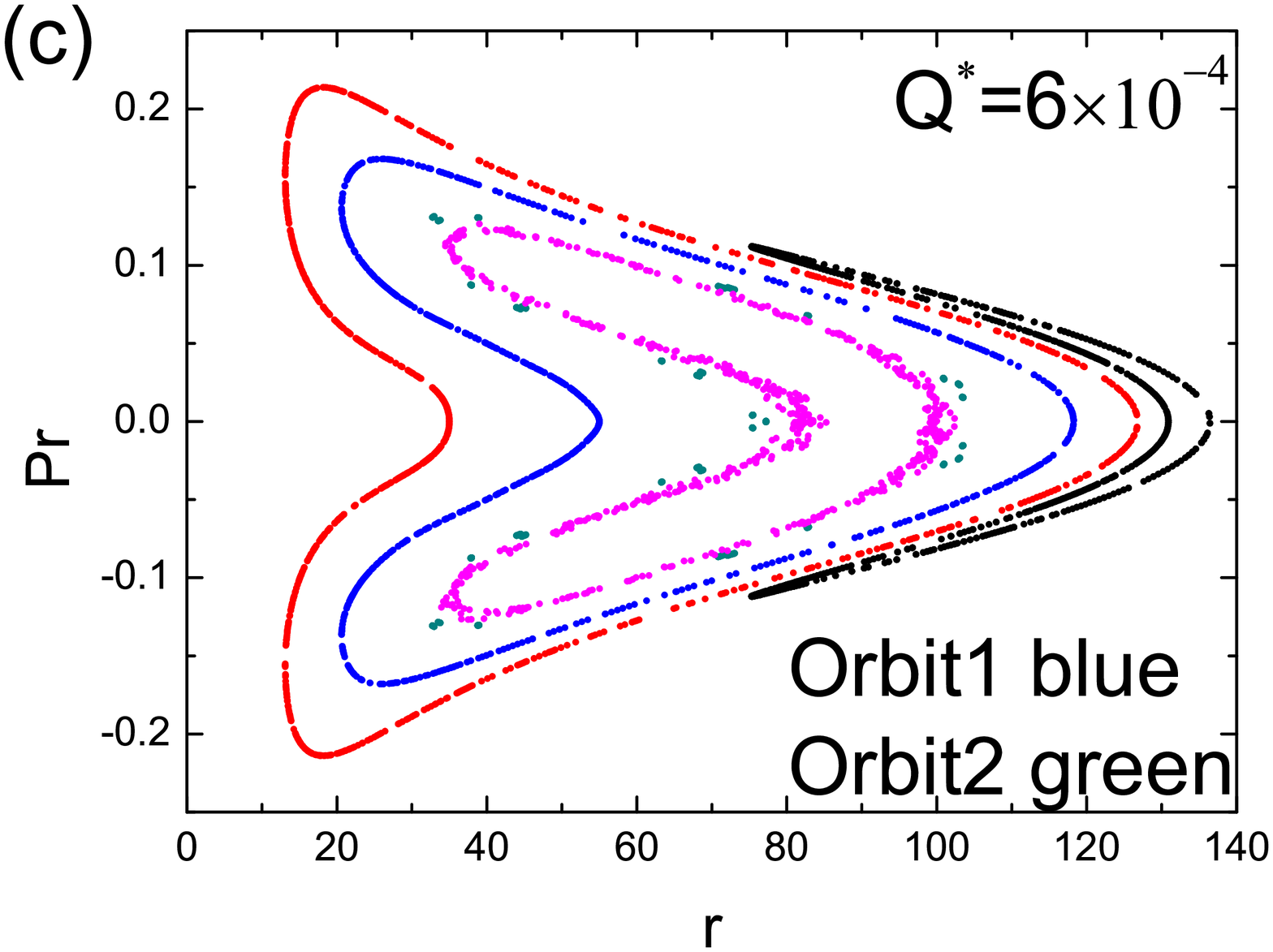}
\includegraphics[width=15pc]{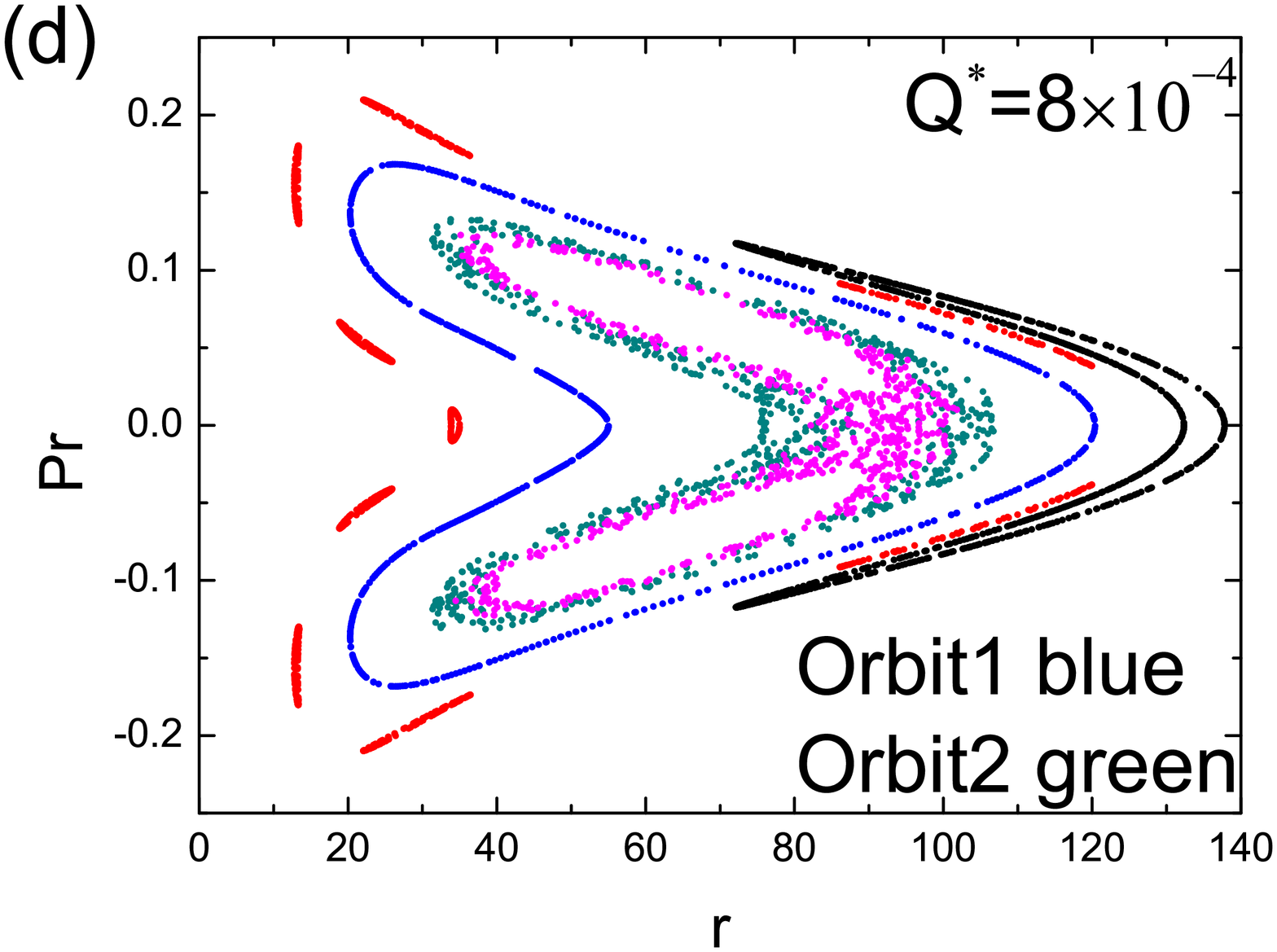}
\caption{Same as Fig. 5(c) but the inductive charge parameters
$Q^{*}$ are different. There is no chaos in (a) and (b). Chaos
occurs for the pink orbit with the initial separation $r=100$ in
(c). Orbit 2 and the pink orbit with the initial separation
$r=100$ are chaotic in (d). }\label{figure6}}
\end{figure*}

\clearpage

\begin{figure*}[!ht]
\centering{
\includegraphics[width=12pc]{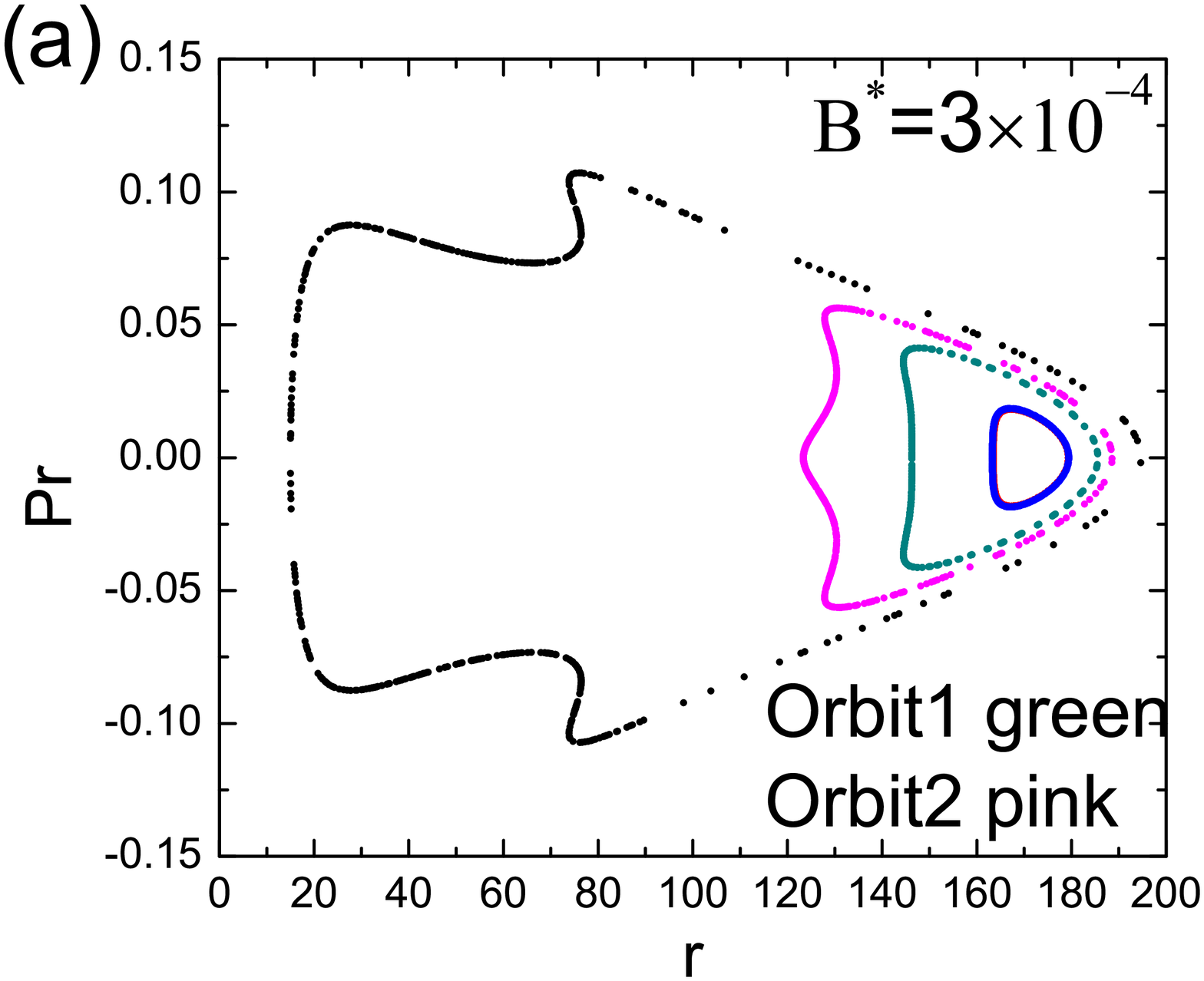}
\includegraphics[width=12pc]{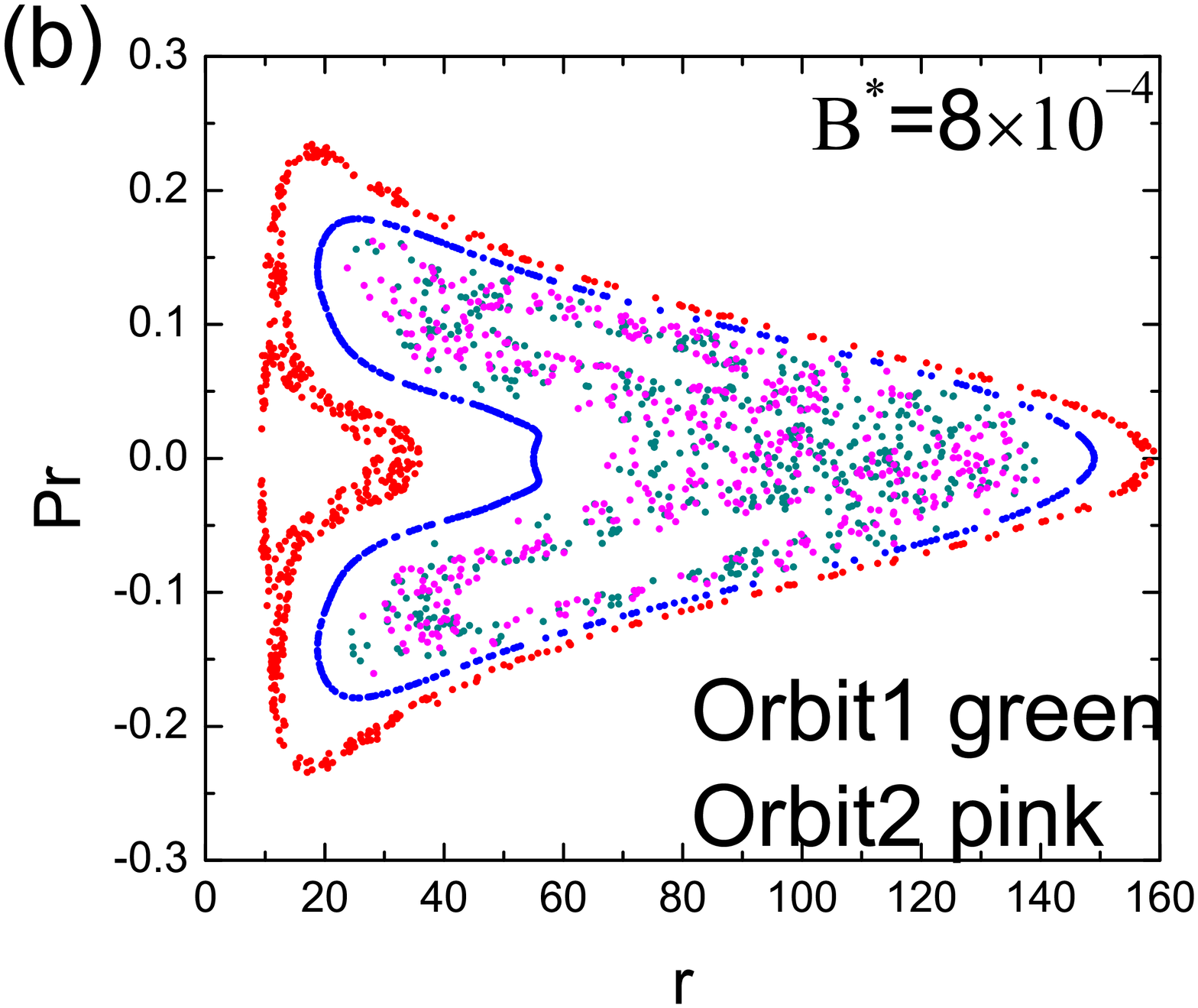}
\includegraphics[width=12pc]{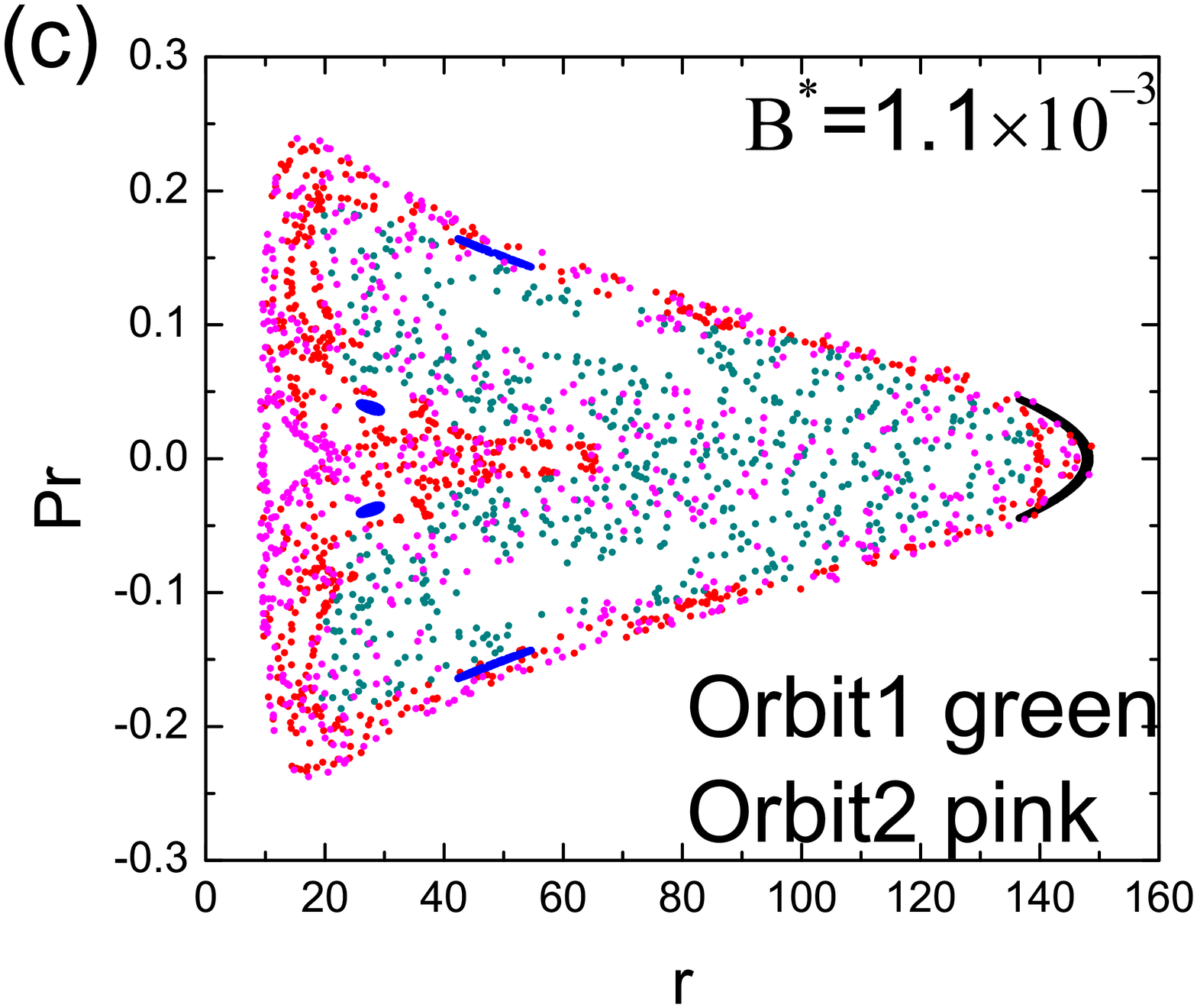}
\caption{Poincar\'{e} sections. The parameters are $E=0.995$,
$L=4.6$, $a=0.5$ and $Q^\ast=0.0003$. Orbits 1 and 2 have their
initial separations $r=75$ and $r=100$, respectively. The initial
separations are $r=15$ for the black orbit, $r=35$ for the red
orbit, and $r=55$ for the blue orbit.}\label{figure7}}
\end{figure*}

\begin{figure*}[!ht]
\centering{
\includegraphics[width=12pc]{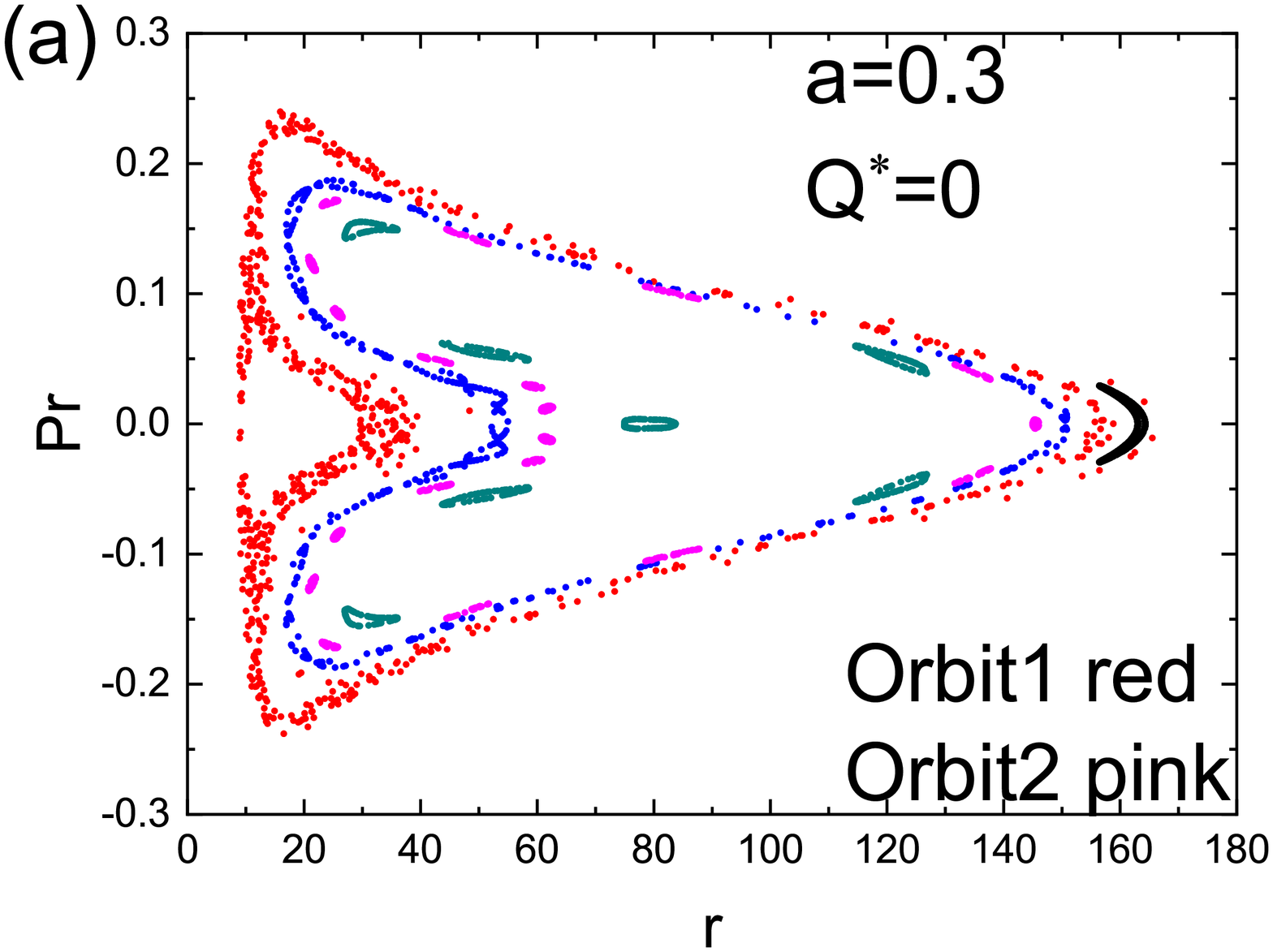}
\includegraphics[width=12pc]{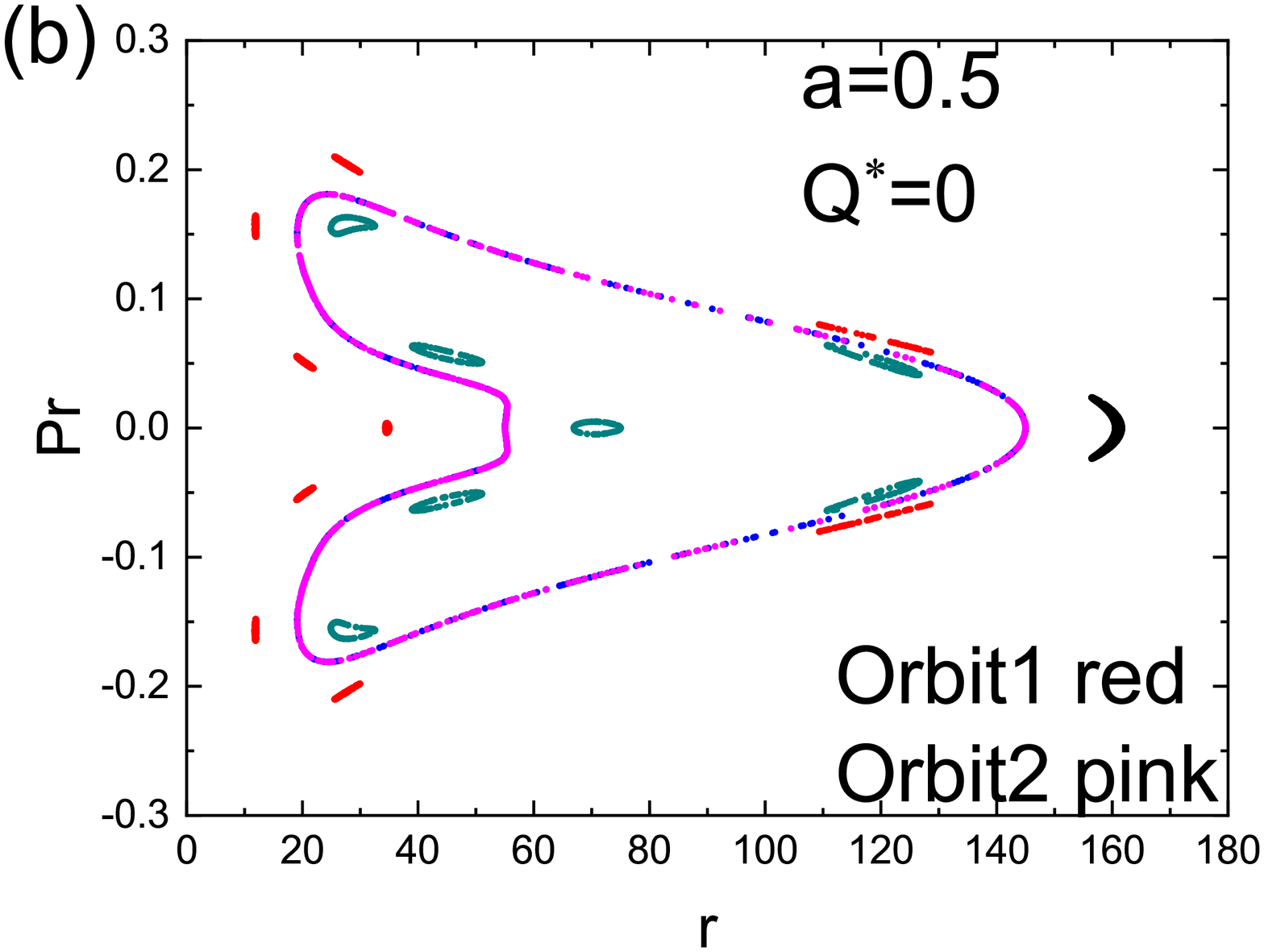}
\includegraphics[width=12pc]{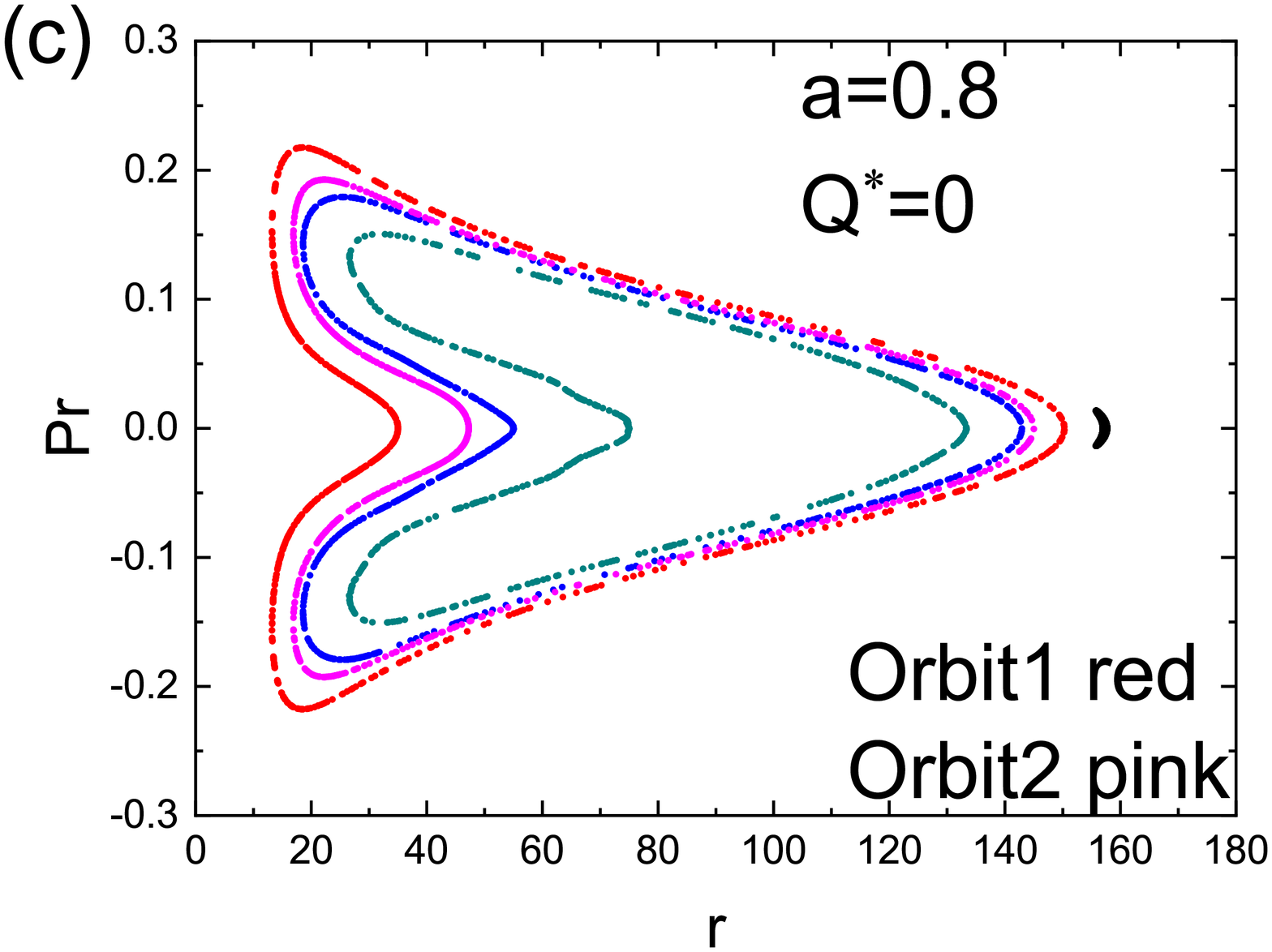}
\includegraphics[width=12pc]{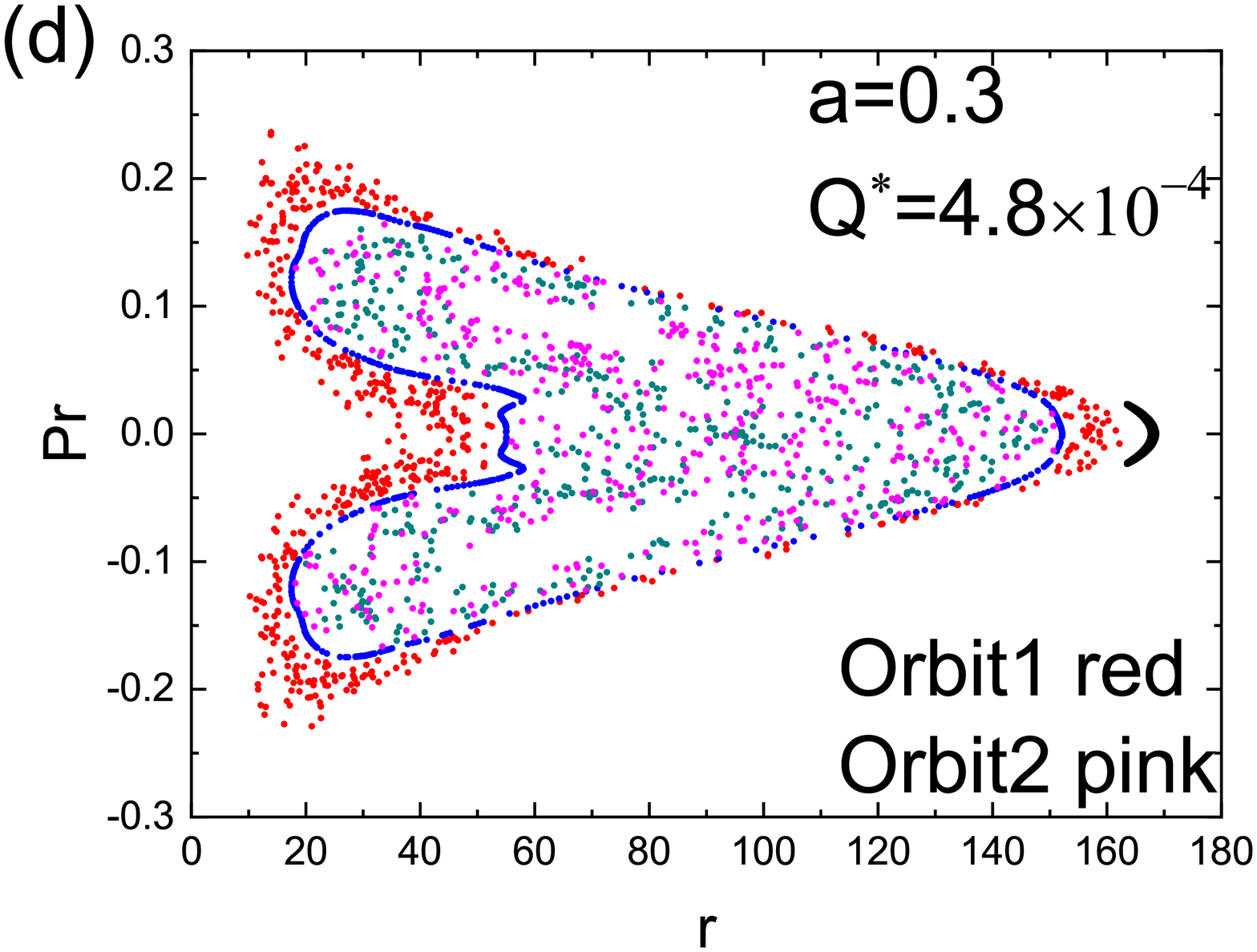}
\includegraphics[width=12pc]{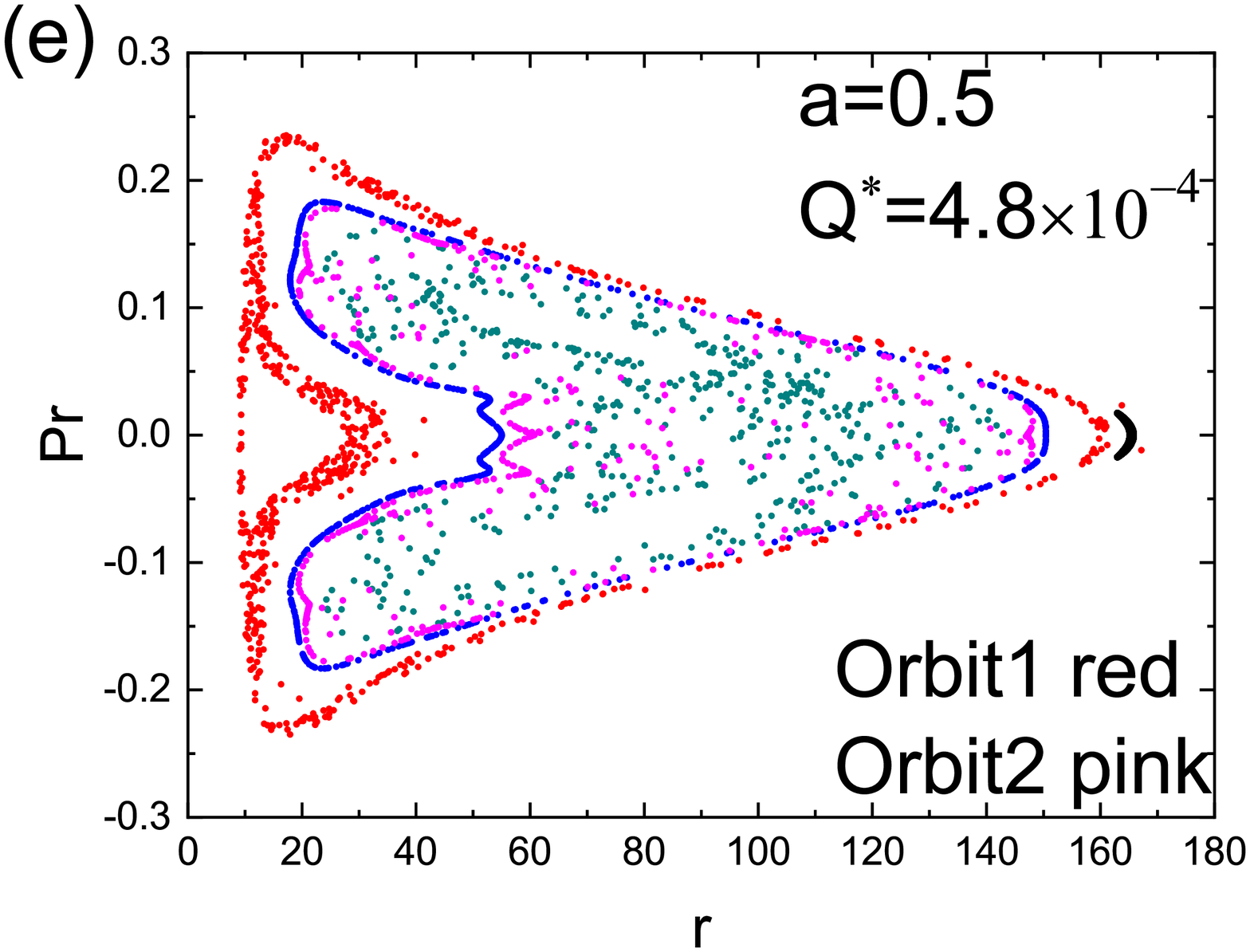}
\includegraphics[width=12pc]{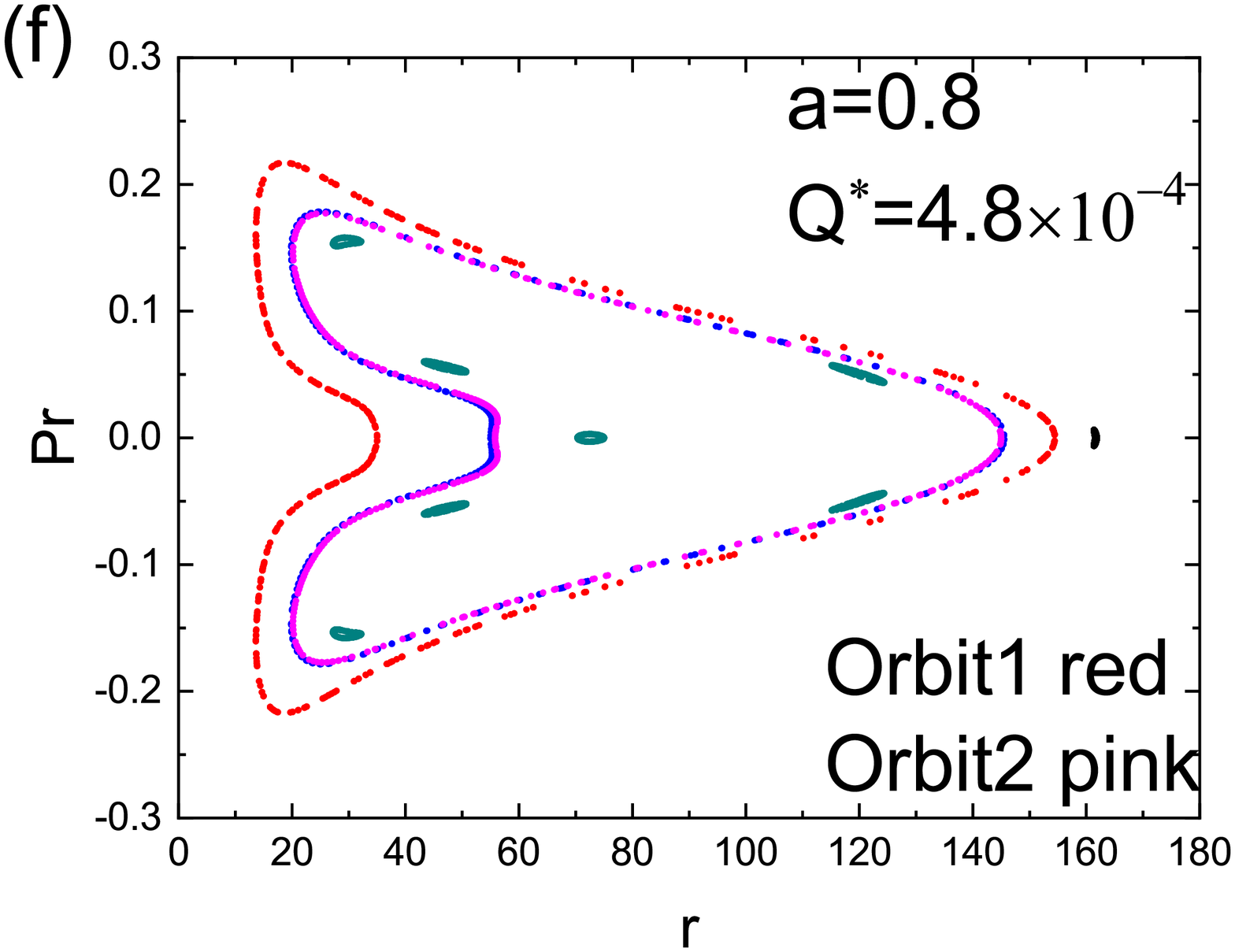}
\caption{Poincar\'{e} sections. The parameters are $E=0.995$,
$L=4.6$ and $B^\ast=0.0008$. Orbits 1 and 2 have their initial
separations $r=35$ and $r=145$, respectively. The initial
separations are $r=15$ for the black orbit, $r=55$ for the blue
orbit and $r=75$ for the green orbit.} \label{figure8}}
\end{figure*}

\begin{figure*}[!ht]
\centering{
\includegraphics[width=12pc]{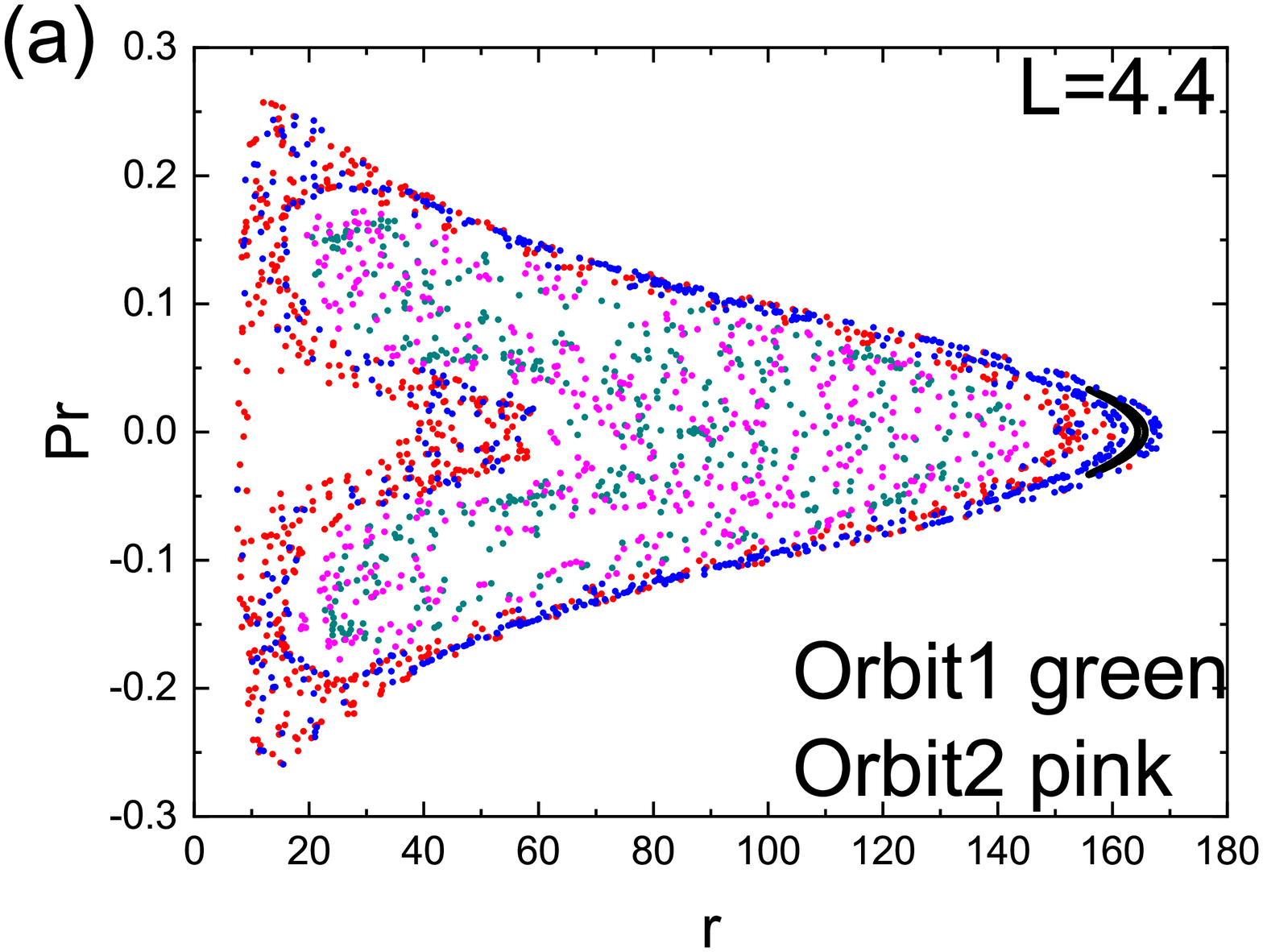}
\includegraphics[width=12pc]{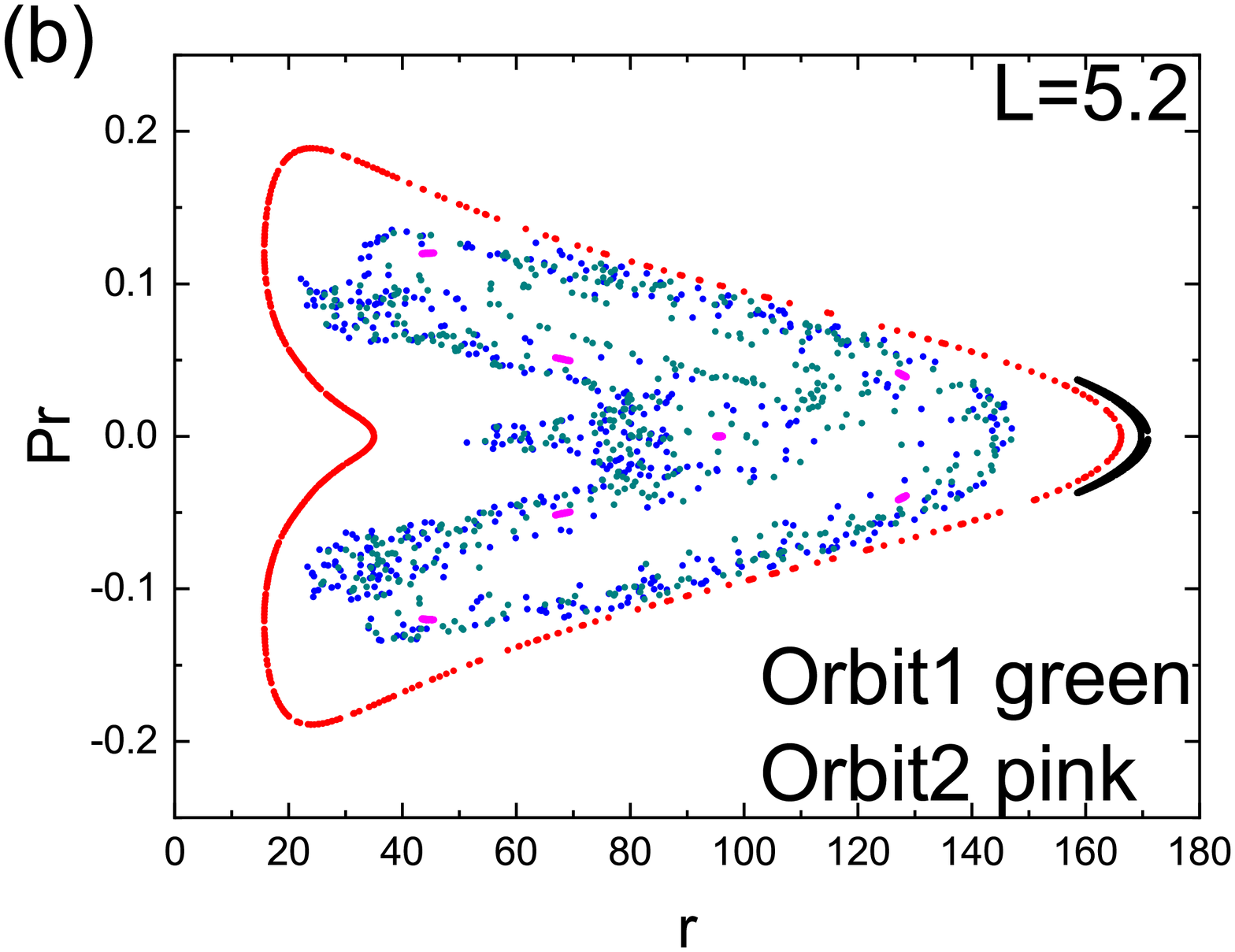}
\includegraphics[width=12pc]{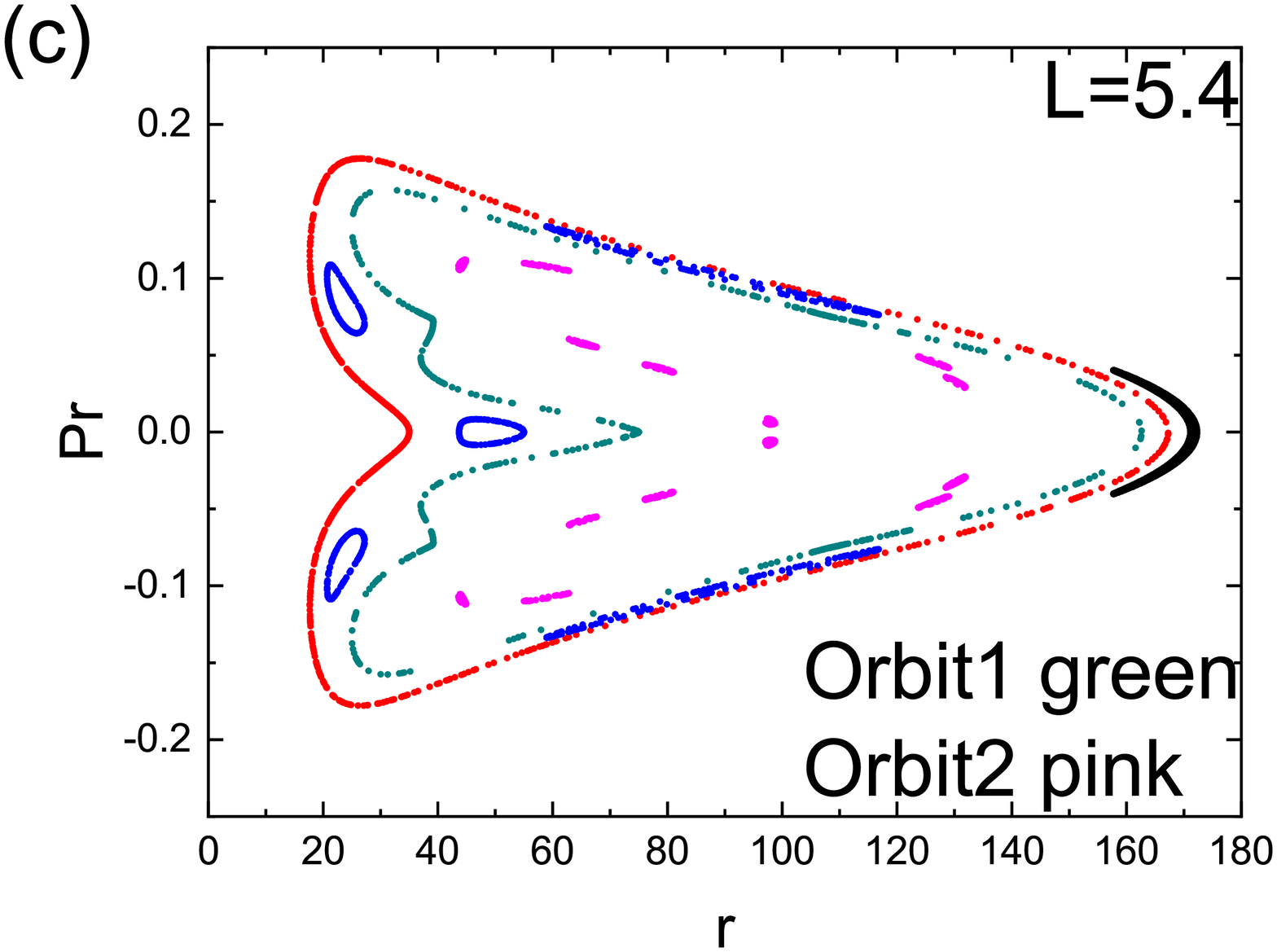}
\caption{Poincar\'{e} sections. The parameters are $E=0.995$,
$a=0.1$,  $B^\ast=8\times10^{-4}$  and $Q^\ast=1\times10^{-5}$.
Orbits 1 and 2 have their initial separations $r=75$ and $r=95$,
respectively. The initial separations are $r=15$ for the black
orbit, $r=35$ for the red orbit and $r=55$ for the blue
orbit.}\label{figure9}}
\end{figure*}

\begin{figure*}[!ht]
\centering{
\includegraphics[width=15pc]{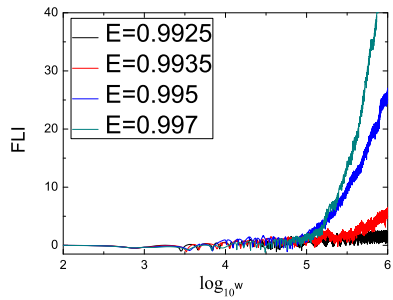}
\caption{Fast Lyapunov indicators (FLIs). The parameters are
$L=4.6$, $a=0.5$,  $B^\ast=0.001$  and $Q^\ast=0.001$. The initial
separation is $r=75$. }\label{figure10}}
\end{figure*}

\begin{figure*}[!ht]
\center{
\includegraphics[width=12pc]{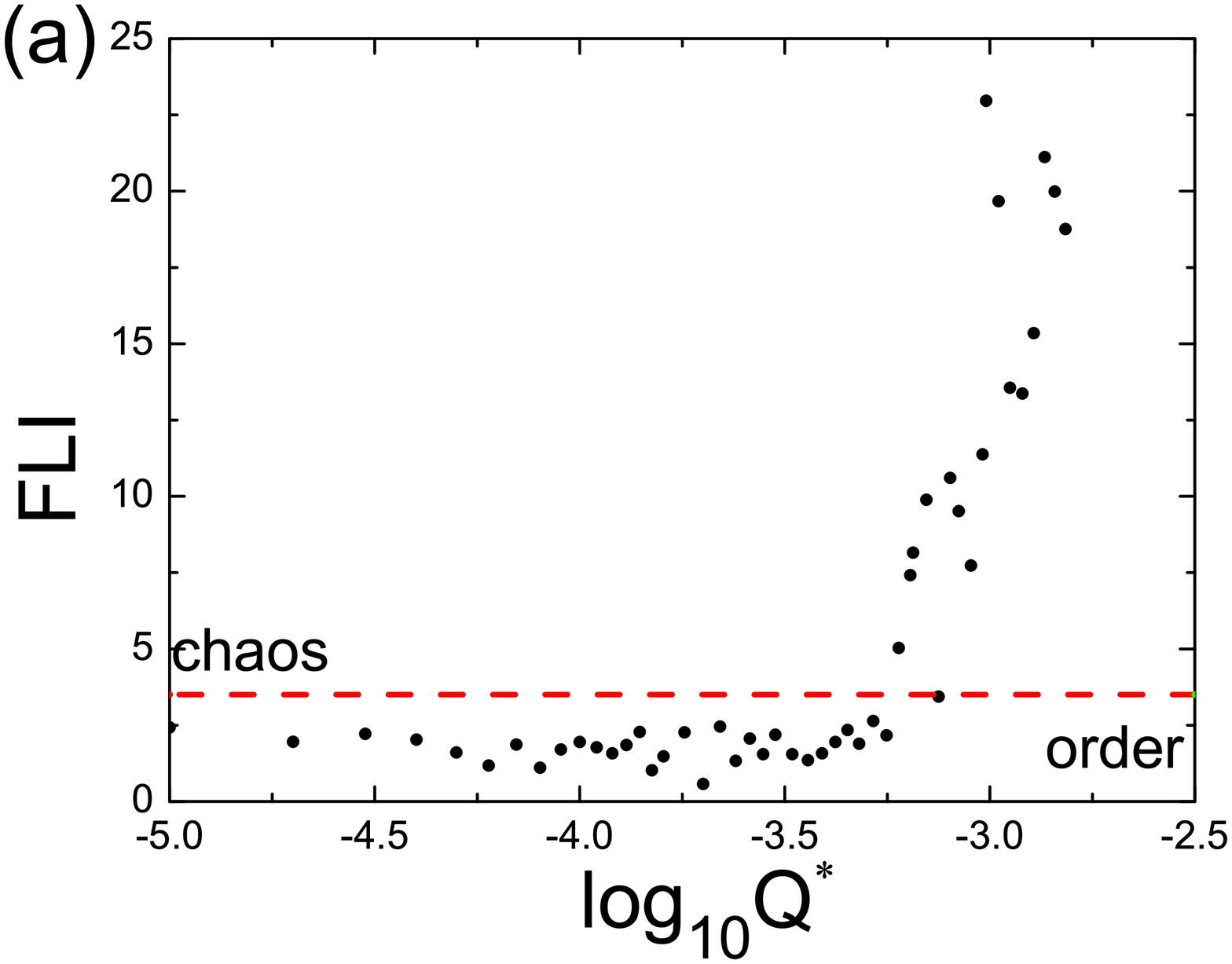}
\includegraphics[width=12pc]{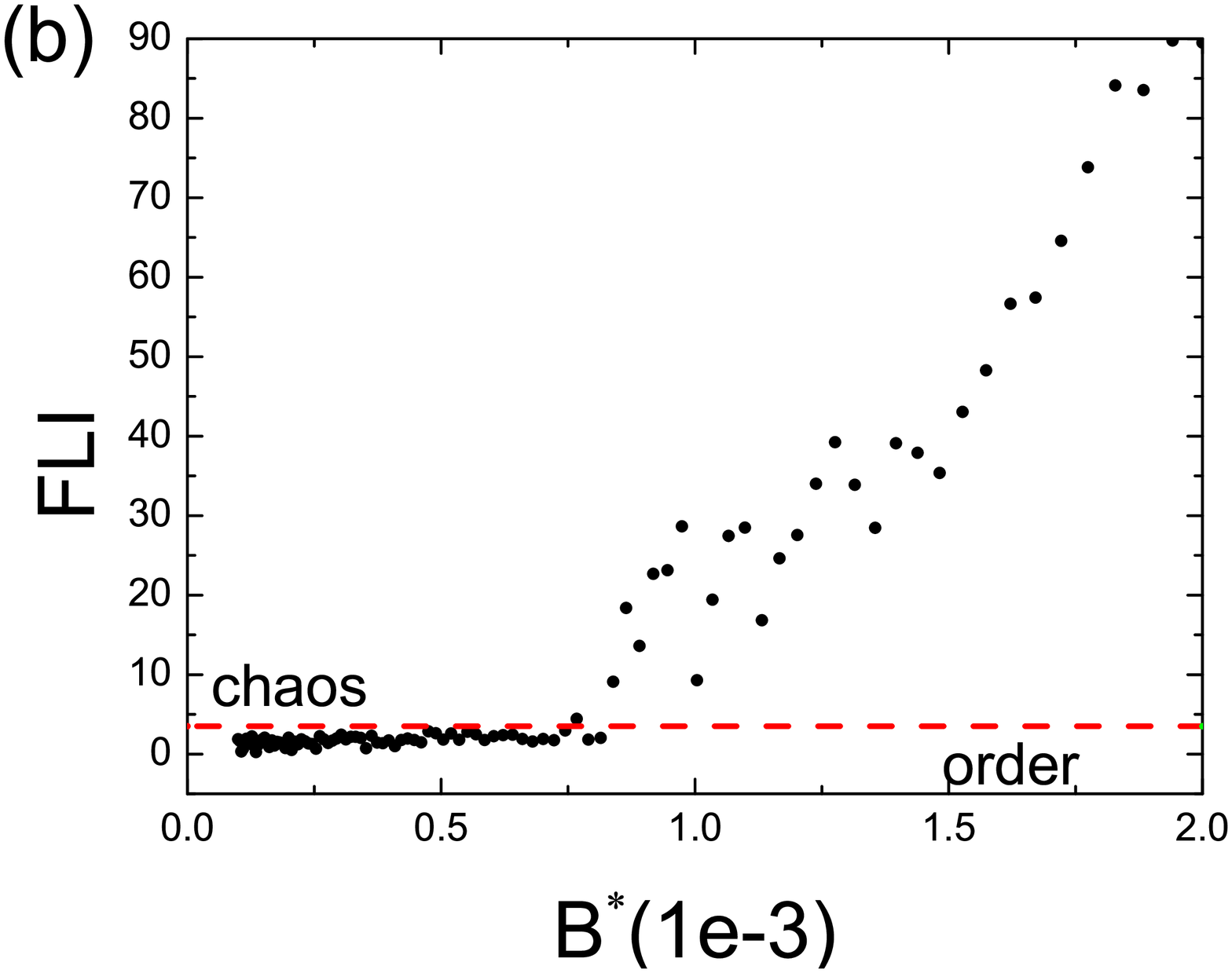}
\includegraphics[width=12pc]{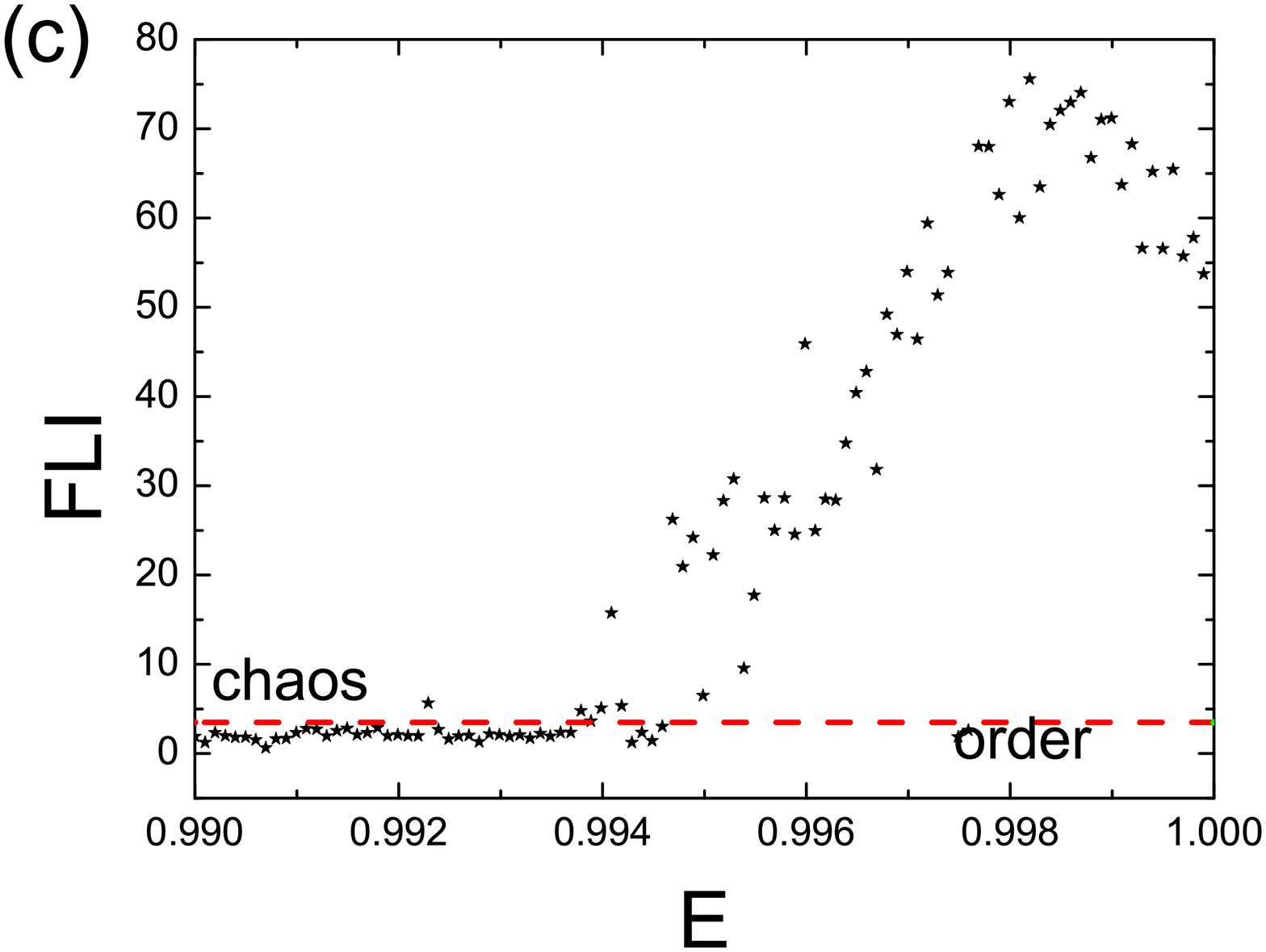}
\includegraphics[width=12pc]{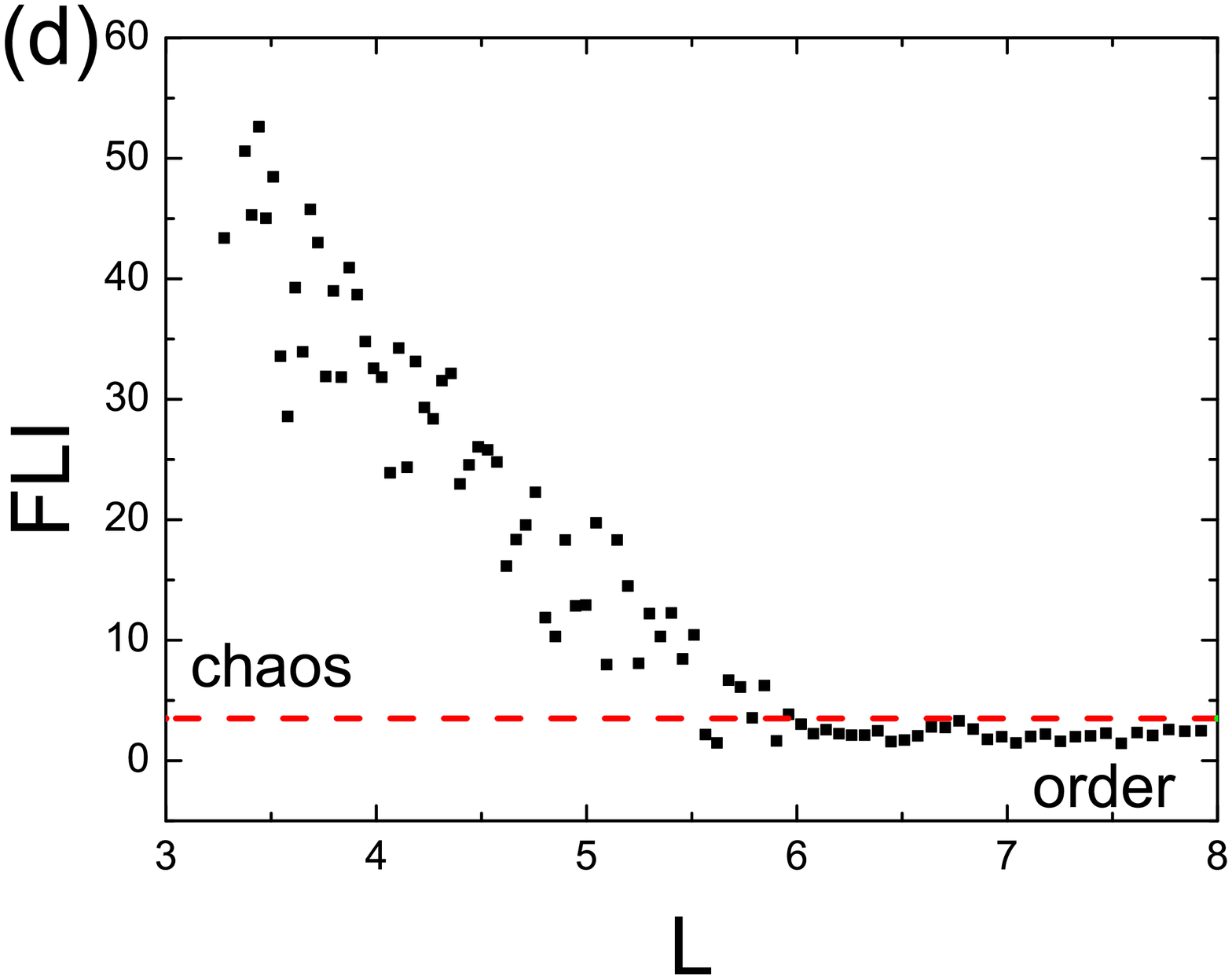}
\includegraphics[width=12pc]{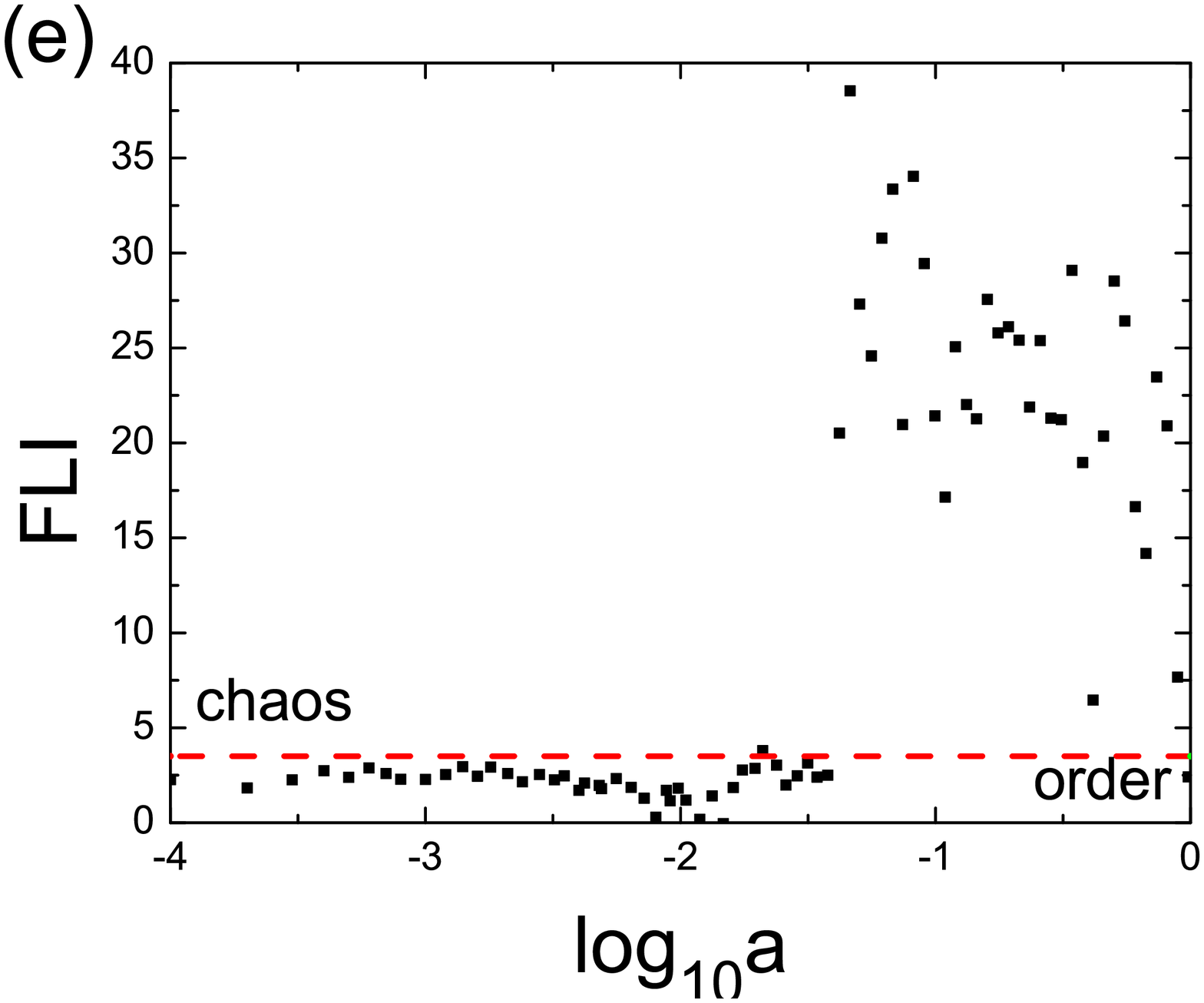}
\caption{Dependence of FLIs on the parameters. Each of the FLIs is
obtained after the integration time $w=10^{6}$. The parameters and
initial separations are as follows. (a) $E=0.9935$, $L=4.6$,
$a=0.5$, $B^\ast=0.001$ and $r=100$.  (b) $E=0.995$, $L=4.6$,
$a=0.5$,  $Q^\ast=0.0001$ and $r=75$. (c) $L=4.6$, $a=0.5$,
 $B^\ast=1\times10^{-3}$, $Q^\ast=1.25\times10^{-4}$ and $r=75$. (d) $E=0.995$,
$a=0.5$,  $B^\ast=1\times10^{-3}$, $Q^\ast=1.25\times10^{-4}$ and
$r=75$. (e) $E=0.995$, $L=4.6$, $B^\ast=1\times10^{-3}$,
$Q^\ast=2\times10^{-7}$ and $r=75$.
 }\label{fig11}}
\end{figure*}

\end{document}